\documentclass{article}

\usepackage{PRIMEarxiv}
\usepackage[utf8]{inputenc}
\usepackage{natbib}
\usepackage{todonotes}
\usepackage{caption}
\usepackage{subcaption}
\usepackage{amsmath}
\usepackage{amsfonts}
\usepackage{amssymb}
\usepackage{float}
\usepackage{alphabeta}
\usepackage{CJKutf8}
\usepackage{verse}
\usepackage{graphicx}
\usepackage{enumerate}
\usepackage{xcolor}

\usepackage{prodint}
\usepackage[printonlyused,withpage]{acronym}
\usepackage{hyperref}
\usepackage{chemfig}
\usepackage{longtable}
\usepackage{emptypage}
\usepackage{lscape}
\usepackage{longtable}
\usepackage{color, colortbl}

 \title{Two-step interpretable modeling of Intensive Care Acquired Infections}
 \author{G. Lancia$^1$,\\
Mathematical Institute,
Utrecht University\\ 
Budapestlaan 6, 3584 CD, Utrecht, The Netherlands\\
\texttt{g.lancia@uu.nl}\\
\And M. Varkila,\\
Julius Center for Health Sciences and Primary Care, University Medical Center Utrecht, The Netherlands,\\
  Department of Intensive Care Medicine, University Medical Center Utrecht, The Netherlands
\And O. Cremer,\\
Julius Center for Health Sciences and Primary Care, University Medical Center Utrecht, The Netherlands,\\
 Department of Intensive Care Medicine, University Medical Center Utrecht, The Netherlands
\And  C. Spitoni,\\
Mathematical Institute, Utrecht University, The Netherlands\\
 \texttt{C.Spitoni@uu.nl} 
}

\begin{document}
\maketitle
\begin{abstract}
We present a novel methodology for integrating \emph{high resolution} longitudinal data with the dynamic prediction capabilities of survival models. The aim is two-fold: to improve the predictive power while maintaining the interpretability of the models. 
To go beyond the \emph{black box} paradigm of artificial neural networks, we propose a parsimonious and robust semi-parametric approach (i.e., a landmarking competing risks model) that combines routinely collected \emph{low-resolution} data with predictive features extracted from a convolutional neural network, that was trained on \emph{high resolution} time-dependent information. We then use \emph{saliency maps} to analyze and explain the extra predictive power of this model.
To illustrate our methodology, we focus on healthcare-associated infections in patients admitted to an intensive care unit.
\end{abstract}


\footnotetext[1]{Present address: Mathematics Department, University of Genoa\\ Via Dodecaneso 35, 16461, Genoa, Italy}


\newpage
\section{Introduction}
\label{sec:intro}

Artificial Neural Networks (ANNs) are very accurate predicting tools when compared to more conventional survival models \cite[]{topol2019high, zeng2022deep, ivanov2022accurate}.
However, they are often seen as \emph{black boxes}, since often it is not possible to express the connection between ANN predictions and input data in a closed form.

ANN models are indeed very difficult to interpret and it is challenging to identify which predictors are the most relevant \cite[]{May11}.  
In contrast, semi-parametric hazard-based survival models (\cite{bible}) are examples of interpretable models, whose hazards can measure (directly or indirectly) the effect of each covariate on the outcome of interest. 

In order to properly model the temporal evolution of the survival process, including longitudinal information (e.g., biomarkers, health status, clinical measurements) as time-dependent covariates is often informative. These covariates are usually \emph{internal} and they require extra modeling to predict survival functions accurately \cite[]{cortese2010competing}.  
The use of Joint Modeling (JM), which attempts to jointly model the longitudinal covariates and the event time, might be then a natural choice \cite[]{proust2009development, rizopoulos2011dynamic, rizopoulos2012joint}. 
Although JMs  can efficiently estimate the underlying parameters when the model is correctly specified, they are sensitive to misspecification of the longitudinal trajectory \cite[]{miss} and they are complex to estimate.

For these reasons, we consider a Landmarking (LM) approach for the dynamic prediction of the outcome of interest (e.g., intensive care unit acquired infections). 
LM is indeed a pragmatic approach that avoids specifying a model for the longitudinal covariates and it is robust under misspecification of the longitudinal processes \cite[]{van2007dynamic, van2011dynamic}. 
The main idea behind LM is to select a point in time $s$ known as a landmark. 
By selecting subjects at risk at $s$  (i.e., left-truncation at time $s$) and by  imposing  administrative right-censoring at time $s+w$ (\emph{horizon time}),  a landmark dataset is then constructed. 
Thus, for a time-dependent covariate $Z(t)$, only  the value $Z(s)$ at
$s$ is considered so that the resulting LM dataset can be analyzed by using standard methods: $Z(s)$ is indeed treated as a time constant covariate. 
In case of competing events, the LM approach can be generalized to the Competing Risks model (LM-CR), see  \cite{nicolaie2013dynamic}. 

The novelty of the manuscript is the inclusion in the LM-CR model of time-dependent information coming from \emph{high-resolution} Electronic Health Record (EHR) data: vital signals recorded in the Intensive Care Unit (ICU) monitors and sampled every minute (i.e., heart rate, mean arterial blood pressure, pulse pressure, arterial oxygen saturation, and respiratory rate). 
A type of deep neural network, a Convolutional Neural Network (CNN), that looks for predicting patterns present in the signals prior to the landmark time $s$, is used as a features' extractor to be included in the main LM-CR model. We hypothesize indeed that these patterns represent additional information, not contained in the \emph{lower-resolution} covariates. 

Although the LM-CR is in itself an interpretable model, we would like to interpret the additional predicting power of the CNN score in terms of the medical conditions of the patients. 
Thus, we studied the pattern recognition performed by the CNN and made it interpretable via a Saliency Map Order Equivalent (SMOE) scale \cite[]{mundhenk2019efficient}; an algorithm that describes the statistics of the activated feature maps of the hidden layers of the network. 
By the SMOE scale, we could visualize the regions of the input data with the highest \emph{saliency} for the prediction. 
Hence, we extracted subsets of the signal with the highest cumulative saliency, to perform a data-driven clustering of patients who are more likely to experience the outcome in the fore-coming prediction window. This approach represents a proof of concept for future applications of our method.

In order to illustrate the methodology, we focused on healthcare-associated infections in patients admitted to an ICU, where they were a major cause of morbidity and mortality \cite[]{mortalityICU, maki2008nosocomial}. Therefore, early identification of infectious events could help physicians in the prevention and management of infectious complications in the ICU \cite[]{dantes2018combatting}.  
Moreover, the dynamic prediction of nosocomial infections is a modeling challenging task. 
The establishment of the presence of infection is not straightforward, and the exact time of infection onset cannot be directly observed. 
Hence, a method that can predict an approaching infection, might give the partitioners valuable lead time to intervene.


The structure of the paper is the following. In Section~\ref{data} we describe the data and define the outcome we want to predict; in Section~3 we introduce the two-step modeling approach; in Section~4 we explain the  design of the CNN, its  training, and the \emph{risk score}'s extraction.  In Section~5 we define and fit the LM-CR model with the inclusion of the  \emph{risk score} extracted by the CNN. Finally, in Section~6 we perform a data-driven clustering based on the SMOE scale analysis of the EHR instances. The \emph{Supplementary material} file contains further information about the data, the selection of the design of the CNN, and  a more detailed explanation of the SMOE scale used in the paper. 

\section{The data}
\label{data}
We analyzed data from the Molecular Diagnosis and Risk Stratification of Sepsis (MARS)-cohort \cite[]{klouwenberg2013interobserver}.  We selected patients $>$18 years of age having a length of stay $>$48 hours, who had been admitted to the ICU of one of the participating study centers between 2011 and 2018. In addition, 
we also used high-resolution data streams from vital signs monitors which had been recorded in the hospital information system at a 1-minute resolution.

As the outcome parameter for our primary modeling attempt, we used the onset of the first occurrence of a suspected Intensive Care Unit Acquired Infection (ICU-AI) within a 24-hour time window from the moment of prediction. The time of infection onset was determined by either the start of new empirical antimicrobial treatment or the sampling of blood for culture (subsequently also followed by antibiotic therapy), whichever occurred first.
The dataset thus consisted of 5075 ICU admissions in which 871 first cases of ICU-AIs occurred. Importantly, the incidence of ICU-AI remained relatively constant across ICU stay at a mean rate of 0.04 (SE 0.01) events per day during the first 10 days in ICU. Median time of onset was 5.25 (IQR 3.80-9.45) days following admission.

We selected candidate predictors among several variables based on literature review, a priori consensus of clinical importance, and prevalence in the study population. 
These covariates include both time-fixed variables reflecting the baseline risk of infection, as well as time-dependent data representing the dynamics of the clinical evolution of patients over time, e.g., laboratory values and physiological response and organ function parameters; see Table~1 and Table~2 in  Section~1 of the \emph{Supplementary Material}.

\section{Two-step modeling strategy}\label{icuai:two_step_model}     

This section offers a concise introduction to the methodology we have proposed.
To take advantage of all longitudinal clinical data and to include observations with different temporal resolutions, we designed our model by means of a \emph{two-step} modeling approach. 
Specifically: 
 
    \begin{enumerate}
    \item[\textbf{Step 1:}] 
    We use a CNN to investigate the longitudinal evolution of {EHR} data.
    The {EHR} data are the high-frequency vital signs recorded in the {ICU} monitors which have a sampling frequency of 1 minute.

    
    The CNN is finalized to obtain a \emph{risk score of infection} (or more simply the \emph{risk score} or CNN score), which will be included among the predictors in Step 2.
    The risk score of infection is designed to prospect the occurrence of an infectious episode at any time of the therapy.
    The higher the risk score, the more the clinical risk of an infectious episode to occur in the near future.
    For ease of use, the risk score ranges from 0 to 1.
    It's crucial to emphasize that the \emph{risk score} is derived by processing information coming from the EHR only.
    Despite achieving values from 0 to 1, the risk score does not represent the probability of infection.
    From a theoretical perspective, the CNN output is not a probability.
    More details about this step are discussed throughout Section \ref{icuai: cnn_at_work}
    
      
    \item[\textbf{Step 2}:] 
    The LM-CR model is fitted, including all explanatory variables, i.e.,  \emph{baseline covariates} (e.g., sex, age, {ICU} admission type, and admission comorbidities), the \emph{low-frequency predictors}  (e.g., consciousness score, laboratory measurements, and bacterial colonization) and the \emph{risk score} obtained by the CNN.
    This model comes with the combination of two models: the Landmark approach and the Competing Risk model.
    The Landmark model allows us to predict any onset of a suspected infectious episode at any moment of the therapy, based on data at one previous moment of the ICU stay.
    The Competing Risk is based on the implementation of a Cox proportional hazard model with two failure causes the onset of an acquired infection and the occurrence of one of exclusive events, such as patient death or discharge from the ICU.
    Additional mathematical details and further insight into this step are elaborated in Section \ref{icuai:landmark_model_J_causes}
    \end{enumerate}

In summary, we trained the CNN using EHR data and evaluated the risk score of infection throughout each patient's ICU stay. This score provides concise information regarding the chance of an impending infectious episode, derived solely from vital sign analysis. 
Subsequently, the risk scores were integrated with other explanatory variables available to us. 
The comprehensive set of predictors was then employed to train the LM-CR model, serving as the primary tool for generating dynamic predictions concerning the onset of infectious episodes.

\section{Step 1: CNN at work}\label{icuai: cnn_at_work}

This section provides information about the CNN model, including its structure, the data it uses, and how it was trained and tested. It also includes a brief explanation of how the scores generated by the CNN were evaluated.

\subsection{Selection of high-frequency instances and imputation}\label{icuai:ehr}
With the term \emph{high-frequency} covariates, we refer to the five high-frequency vital signs available to us, namely Heart Rate (HR), mean Arterial Blood Pressure (ABP), pulse pressure (PP), functional oxygen saturation (\chemfig{SaO2}), and Respiratory Rate (RR). 
As mentioned, these predictors are sampled with a sampling frequency equal to one minute. 
These data were arranged in various 24-hour time series (i.e., each time series contains 1440 records, one record per minute). 

Thus, we selected and extracted the \emph{time series instances} as follows:
\begin{enumerate}

\item We excluded the final 24 hours of data for patients who passed away during their ICU stay. 
This was done to eliminate time windows that might contain unrepresentative information, such as extreme or abnormal records resulting from medical decisions to withhold treatment in the last 24 hours before death. 
Hence, by using these records the CNN could be biased, and the classification task would be made tougher.


\item Starting from admission time $\tau^i_0$ of the $i$-th patient, we partitioned all physiological vital signs into time windows of width $w= 24$  hours until achieving the final time $T^i_\ell$ of the patient record (defined as in point 1 for the patients who died during the stay). 
Therefore, we obtained the set of intervals $\mathcal{P}^i$ for the patient $i$: 
$$
\mathcal{P}_i:=\bigcup_{k\ge 1}\left\{[\tau^i_0+(k-1)w,\min(\tau^i_0+k w,T^i_\ell)]\right\}
$$
Likewise, we defined the set of time windows \emph{shifted by $\delta$} as:
$$
\mathcal{P}_i^\delta:=\bigcup_{k\ge 1}\left\{[\tau^i_0+\delta+(k-1)w,\min(\tau^i_0+\delta+ k w,T^i_\ell)]\right\},
$$
provided that $T^i_\ell\ge\tau^i_0+\delta$. 
Hence, the time windows selected for the patient $i$ are the ones belonging to the set $\mathcal{P}_i^{\textnormal{total}}:=\mathcal{P}_i\cup \mathcal{P}_i^{8\textnormal{hrs}}\cup \mathcal{P}_i^{16\textnormal{hrs}}$; see Figure~\ref{fig:chunks}.
\begin{figure}
    \centering
    \includegraphics[width= \textwidth]{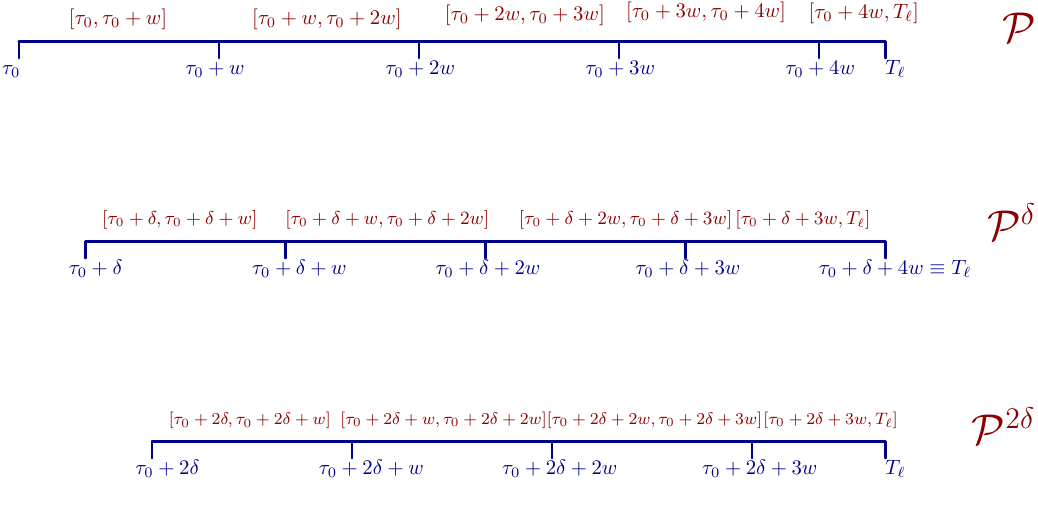}
    \caption{Example of time windows selected for one patient. $\tau_0$ denotes the admission time of the patient, while $w$ the amplitude of the prediction window. The set with all these prediction windows is denoted with $\mathcal{P}$. The picture below shows the selection of windows shifted by a quantity $\delta$; the set of windows is denoted with $\mathcal{P}^{\delta}$. Similarly, the selection of windows shifted by a quantity $2\delta$ is also shown}
    \label{fig:chunks}
\end{figure}
The collection of the time windows in $\mathcal{P}_i^{\textnormal{total}}$ (i.e., consecutive windows of 24 hours and their translations of 8 and 16 hours), allows chunk the longitudinal evolution of the signals coherently with the way we extracted the low-frequency time-dependent covariates of Step 2.
We shall refer to the portion of the five vital signs signals corresponding to an interval in $\mathcal{P}_i^{\textnormal{total}}$ with the term \emph{time series instance}.
\begin{figure}
    \centering
    \includegraphics[width= \textwidth]{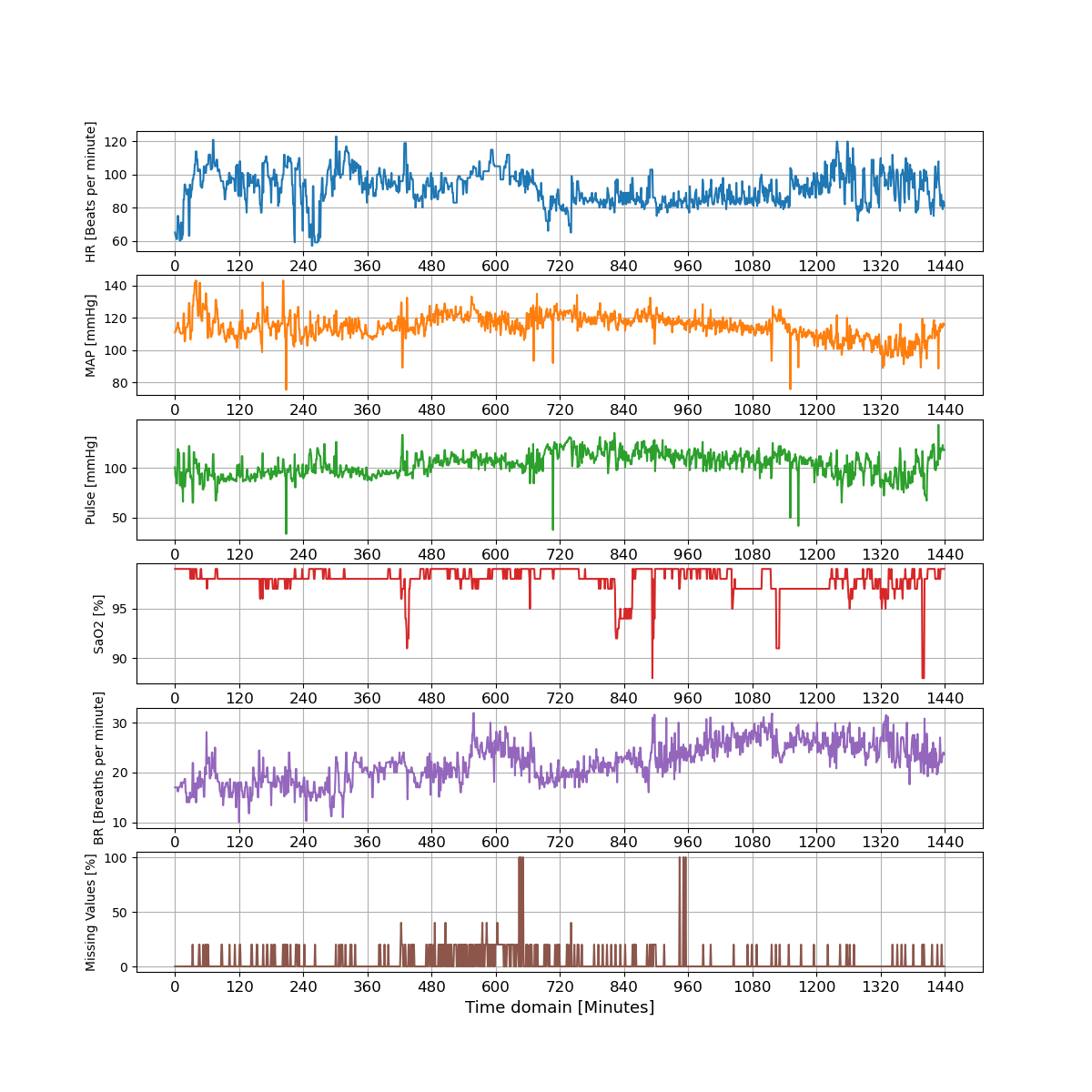}
    \caption{Example of \emph{time series instance}. x-axis: time-domain (24 hours). y-axis: the values taken by each time series feature. 
    In specific, HR in blue, ABP in orange, pulse pressure in green, $\operatorname{SaO_2}$ in red, BR in purple, and the auxiliary time series (with the missing values incidence) in brown.}
    \label{fig:ICUAI_example_instance}
\end{figure}

\item Per each patient $i$ who has acquired no infection during his/her stay in the {ICU}, we termed his/her time series instances as the \emph{not-infected} instances. 
For such a patient we considered all time series instances whose time windows are in $\mathcal{P}_i^{\textnormal{total}}$.

\item  For each patient $i$ who acquired an infection during the stay in the ICU, we first divided his/her ICU stay as in point~2 ($\mathcal{P}_{i}^{\text{total}}$).
We then labeled as an event all the time windows where an ICU-AI event has occurred (i.e., those time windows including the time-stamp at which the ICU-AI episode was recorded).
Likewise, we also labeled all the time windows preceding the time window containing the onset  of an ICU-AI event as outcome events. 
By doing this, we tagged as an event those time series instances anticipating at the most the 24 hours prior to the moment when the ICU-AI episode was reported.  
This choice comes naturally with the necessity of modeling the high volatility of the outcome variable under the exam.
Unlike other events,  such as death in the ICU, the onset of acquired infections cannot be detected at one precise moment unless the worsening of clinical conditions has become overt.
All remaining time windows associated with no infectious episode, are therefore treated as not-infected.
\item We considered the first ICU-AI episode while discarding all the other recurrent episodes from the same patient.  
More precisely, all the instances following the first infection were discarded. 
\item We equipped each \emph{time series instance} with an extra time series monitoring the presence of missing values: This strategy allowed us to track the percentage of missing records at each time stamp.  
\end{enumerate}

Hence, each \emph{time series instance} was described by a $6\times 1440$ matrix, whose rows represent the type of \emph{time series features} (i.e., HR, ABP, pulse pressure, \chemfig{SaO2}, BR and missing records) and the columns the time domain (note that 1440 corresponds to the total number of records in a day; calculated as 24 hours multiplied by 60 minutes per hour)
The illustration of one sample \emph{time series instance} is shown in Figure~\ref{fig:ICUAI_example_instance}. 

Missing values of EHR have been imputed by using a zero-order spline, i.e., the Last Occurrence Carried Forward (LOCF) method. 
Despite being a very simplistic approach, we noted that it has already been applied in some other similar contexts; for example \cite{gandin2021interpretability, deng2023dynamic}.
In our case, however, the simplicity of this imputation method is mitigated by the inclusion, per each time series instance, of an extra time series reporting the intervals and the number of vital signs that are missing.
This strategy helps the CNN model to better recognize the correct informativeness of patterns, that are transmitted through the first layers.
Generally speaking, the first convolutional layer is responsible for processing the features of complete vital signs; when it occurs the extra time series is silent and zero-valued.
However, when processing missing signals, the processing of the extra time series acts like an extra term readapting the argument of the activation function. 
This adaption is mutated by some proper weights that are sharpened during the learning phase.

In addition to theoretical considerations, the choice of the LOCF imputation method was also motivated by a comparative analysis involving two alternative methods.
The first method employed was the multivariate kNN (k Nearest Neighbours)\cite[]{troyanskaya2001missing}, originally designed for matrix data but adapted for time-series instances in our study. 
The other one, hereafter referred to as the \emph{ICUAI-Imputation} method, was tailored to better align with the inherent nature of missing values in the ICU context. 
In essence, missing intervals with an amplitude exceeding 4 hours were substituted with a constant out-of-range value (e.g., 100), while intervals shorter than 4 hours were imputed with a null constant value. 
The rationale behind the ICUAI-imputation method drew inspiration from a practical medical perspective in managing Electronic Health Records (EHR). 
Intervals of approximately 4 hours or longer typically corresponded to the duration of surgical operations, while shorter intervals were often associated with the temporary interruption of ICU monitoring, resulting from the unintentional detachment of devices, whether by a patient or due to device malpositioning.
With this method, we systematically filled specific types of missing intervals with designated placeholder values; this way, the imputed patterns of missing intervals could also include distinct clinical events of interest occurring during the ICU stay.
A comparison among the performances of the CNN model adopting all these different imputation methods revealed the LOCF to be the most performing.
Deeper insights into this are available in Section 2 of the \emph{Supplementary Material}.


Before feeding the vital signs into the CNN model, we preprocessed the vital signs. 
In particular, we applied a single linear transformation to all time-series features to condense them into the range [-1, 1]. 
We devised and applied a linear transformation to all time series features of the same kind. 
Thus, considering the overall statistics of available vital signs, we crafted linear mappings for each time-series feature to rescale them within the [-1, 1] domain.
More insight into the pre-processing is available in the Section 2 of the \emph{Supplementary Material}.

We remark that in order to illustrate our methodology, we opted to concentrate on a 24-hour time window primarily.
The analysis was also repeated with a 48-hour window (as done in Section 2 
of the \emph{Supplementary material}). 
However, the larger the prediction window, the larger the dimensionality of the input data.

\subsection{Design of the CNN}\label{icuai:two_step_model_step1}

The last decade has shown how the predicting skill of CNN turned out to be highly successful in solving various tasks in many different contexts, e.g. image recognition \cite[]{liu2018feature, zheng2017learning, lou2020face, kagaya2014food}, anomaly detection \cite[]{kwon2018empirical, naseer2018enhanced, staar2019anomaly}, and time series forecasting \cite[]{borovykh2017conditional, selvin2017stock, livieris2020cnn, guo2019combined} among others.
Indeed, this class of Artificial Neural Networks {(ANNs)} is specifically designed to work with grid-structured data.
Its great ability in processing complex multi-level data is mostly due to the combination of both convolutional and max-pooling operators which allow the encoding of the sequentiality of the patterns along the multi-dimensional domains of input data.
For example, the optimal search of the convolutional filters of the convolutional layer in a one-dimensional CNN has the scope of obtaining the  most linearized latent representation of the input time series instances.
%
\begin{figure}
    \centering
    \includegraphics[width= 8cm]{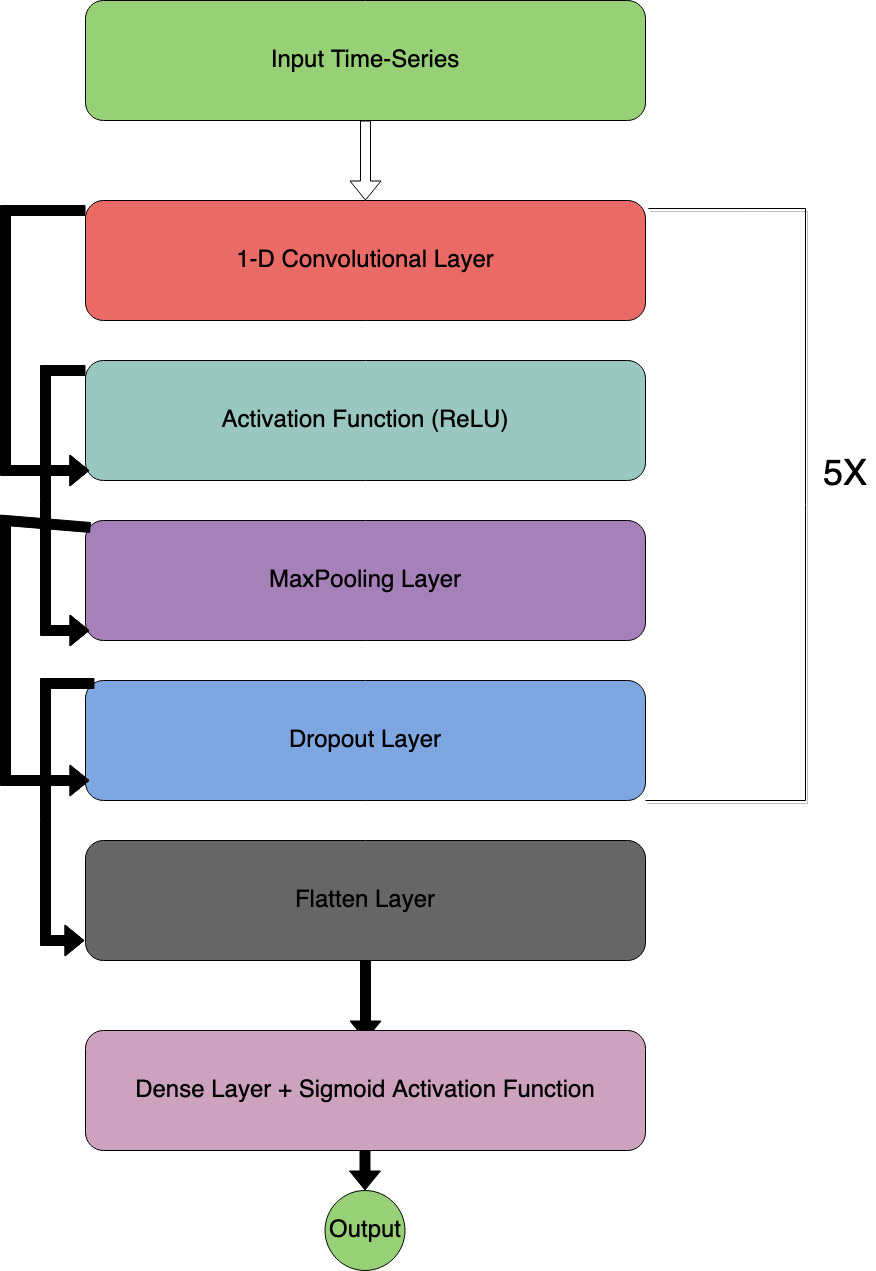}
    \caption{\small Schematic illustration of the CNN model. The input signal is processed by a convolutional layer (128 filters of size 3). 
    The \emph{ReLU} function is applied before a \emph{max-pooling} operator which reduces the size of the features. After each \emph{max-pooling} layer follows a \emph{dropout layer} whose dropout rate is 0.25. 
    This sequence of hidden layers is repeated five times.
    The feature maps are then flattened into an array (flatten layer) and then propagated through a \emph{fully-connected} layer (dense layer) with a sigmoid activation function.} 
    \label{fig:schematic_CNN}
\end{figure}
To evaluate the desired risk scores of infection from EHR, we have therefore chosen to utilize a convolutional network: its architecture is composed of convolutional, pooling, and dense layers only. 
The choice of a CNN seems natural since we are looking for translational invariant patterns that might be present in any sub-interval in the time series. 

In order to give quantitative grounds to this reasoning, we compared CNN's accuracy with other traditional Machine Learning and NN-based models, namely Logistic Regression (LR), linear Supported Vector Machine (SVM), Multi-Layer Perceptron (MLP), and CNN-LSTM networks (where LSTM stands for Long Short-Term Memory).
That is, we trained and validated the mentioned models over a fine grid of hyperparameters.
We introduce here that the performance of models was evaluated using the area Under the Receiver Operating Characteristic curve (AUROC), or more simply AUROC score.
Once again, this metric proved to be a suitable choice for gauging the performance of the models.
The AUROC scores of each model have been listed in Table \ref{tab:models_selection_keypoints}.
However, it's important to outline that in order to ensure the candidate models could robustly investigate longitudinal evolution across different time scales of interest, we examined their performance using both 24-hour and 48-hour prediction windows (i.e., the 24-hour instance and 48-hour instance models of Table \ref{tab:models_selection_keypoints}. 
For the 48-hour instance, the selection was made by readapting the strategy of Section \ref{icuai:ehr}).
The results of Table \ref{tab:models_selection_keypoints}revealed that the CNN model did not emerge as the absolute top performer in predictive accuracy. 
However, we were motivated to select it by the fact that it can model the risk score of infection at best when considering both the 24-hour and 48-hour instances. 
Despite the 24-hour CNN-LSTM model potentially exhibiting slightly higher accuracy than the CNN, we observed that the latter displayed more precise predictive performances, even with 48-hour time-series instances. 
The difference in AUROC between both models is minimal for the 24-hour model but becomes more pronounced for the 48-hour model. 
Ultimately, the CNN model demonstrated a more robust skill in capturing relevant patterns in both contexts.

Further details regarding the model selection strategy are available in Section~2 of the \emph{Supplementary Material.} 


\begin{table}[h]
    \centering
    \begin{tabular}{|l|l|l|}
    \hline
     \textbf{Model} & \textbf{AUROC (24-hour model)} & \textbf{AUROC (48-hour model)}\\
     \hline
      LR  & 0.59 ± 0.01 & 0.59 ± 0.01\\
      SVM & 0.57 ± 0.01 & 0.57 ±0.01\\
      MLP & 0.63 ± 0.01 & 0.61 ± 0.01\\
      CNN & 0.72 ± 0.01 & 0.68 ± 0.02\\
      CNN-LSTM & 0.74 ± 0.01 & 0.59 ± 0.01\\
      \hline
    \end{tabular}
    \caption{Model Selection summary: The highest performance achieved during the validation phase, measured by AUROC, is reported for each investigated model. The columns displaying AUROC scores represent either the 24-hour instance model or the 48-hour instance model. AUROC scores have been rounded to the nearest second decimal. Errors were assessed using the Standard Error Mean, and if too short, they were substituted with the minimum error, i.e., 0.01.}
    \label{tab:models_selection_keypoints}
\end{table}

We opted for a CNN design, due to its accuracy and the possibility of applying the saliency maps analysis, as presented in Section~\ref{icuai:XAI_icufai}.


The final architecture chosen for the CNN is the following:
\begin{enumerate}
    \item \emph{Convolutional Layers}: The number of filters on each layer is 128, and each filter has a size of 3 (pixels).
    We call a \emph{feature map} the output of a filter applied to the previous layer. 

    \item \emph{Activation Layer}: The ReLU function (i.e. $\textnormal{ReLU}(x):= \max(0,x)$) is applied after each convolution operator. 
    This application of a non-linear activation function on the feature maps gives rise to the \emph{activated feature maps}.

    \item \emph{Max-pooling layer}: The activated feature maps are resampled via a max-pooling operator with a pooling size of 2 (sub-sampling).
\end{enumerate}
Also,  a \emph{dropout layer} with a dropout rate of 0.25 is included after each max-pooling layer.
This sequence of hidden layers is repeated five times.
The last feature map is flattened into an array and then propagated into a \emph{fully-connected} layer (dense layer) with a sigmoid activation function. 
The activation function returns a positive output between 0 and 1, that is, the risk score.
The architecture of the chosen CNN is sketched out in Figure~\ref{fig:schematic_CNN}. 

\subsection{Training and overall evaluation of the CNN}
\label{section:tren}

Before training the model with the {EHR} data, we made a selection among the time series instances available to us.
Indeed, we opted for under-sampling the total amount of time series instances. 
This choice has a double reason: We mainly wanted to avoid the great majority of risk score infections coinciding with the CNN outputs obtained during either the training or the validation phase.
In other words, we aimed to incorporate in the Landmark model a risk score that originated from time series instances that were never propagated through the CNN before.
The second reason has simply to do with the intrinsic difficulty of CNN in getting well-trained on very imbalanced datasets.
The number of time series instances in the case group (i.e., those instances representing the {ICU-AI} episodes) was less than one-twentieth of the total amount of time series instances in the control group (i.e., those instances not representing the {ICU-AI} episodes). 
Thus, we opted for fitting the {CNN} model on a population of \emph{time series instances} with a {control-case ratio} of 8:1 (i.e., the number of time series instances in the control group is 8 times larger than the case group).
It is important to remark we applied a random under-sampling on the control group only. 

The fit of the model was designed to optimize the binary cross-entropy loss function through the ADAM algorithm \cite[]{kingma2014adam}.
 Therefore, we trained the {CNN} to solve a binary classification task.
 The difference of AUROC between the Deep-LM-CR and the LM-CR model (i.e., both Landmark models with and without the CNN score, respectively; see Sec. \ref{icuai:Deep-LCPH}) is used to evaluate the relative goodness between the two models.
 In addition to this, we also considered the Brier score \cite[]{brier1950verification} as an alternative metric for assessing the prediction power of the model, while Brier Skill \cite[]{wilks2011statistical} is utilized to assess the relative increase in predictive performance of Deep-LM-CR with respect to the LM-CR.
Although our main interest is not in the prediction formulated by the CNN itself, we also needed to guarantee that the CNN model was able to classify the \emph{time series instances} and encode informative patterns about the impending onset of an ICU-AI. 
Internal validation was performed using the \emph{5-fold cross-validation} method; when validating the performance of CNN models as binary classifiers, the data were split into 5 different folds. 
The overall AUROC is then the average over the 5 folds. 
The search for the optimal configuration was conducted through the validation of the models over a fine grid of hyperparameters. The validation strategy adopted was also the \emph{5-fold cross-validation}.
Additional details are available in Section 2 of \emph{Supplementary Material} in which we delved into the CNN model's AUROC variation of three key hyperparameters

\subsection{CNN Risk score}

The extraction of the CNN score and its inclusion in the LM-CR model represent the novel ideas of the manuscript.
The risk score of infection is evaluated by means of the CNN, whose architecture was discussed in Section~\ref{icuai:two_step_model_step1} and its training phase in Section~\ref{section:tren}.

Thus, the procedure for evaluating the risk scores is the following:
\begin{enumerate}
    \item Consider the vital signs of patient $i$ (HR, ABP, pulse pressure, \chemfig{SaO2}, and RR) and the time series flagging the missing records.

    \item Starting from the ICU admission time, extract the 24-hour \emph{time series instances} by means of an 8-hour sliding time window (see Section~\ref{icuai:ehr}), corresponding to the intervals in $\mathcal{P}_i$.

    \item Propagate the \emph{time series instances} through the hidden layers of the fitted {CNN} model and evaluate the risk score.

    \item Assign the risk score to the corresponding time-stamp (i.e., day-month-hour-minute).
\end{enumerate}
\begin{figure}
    \centering
    \includegraphics[width= 9cm]{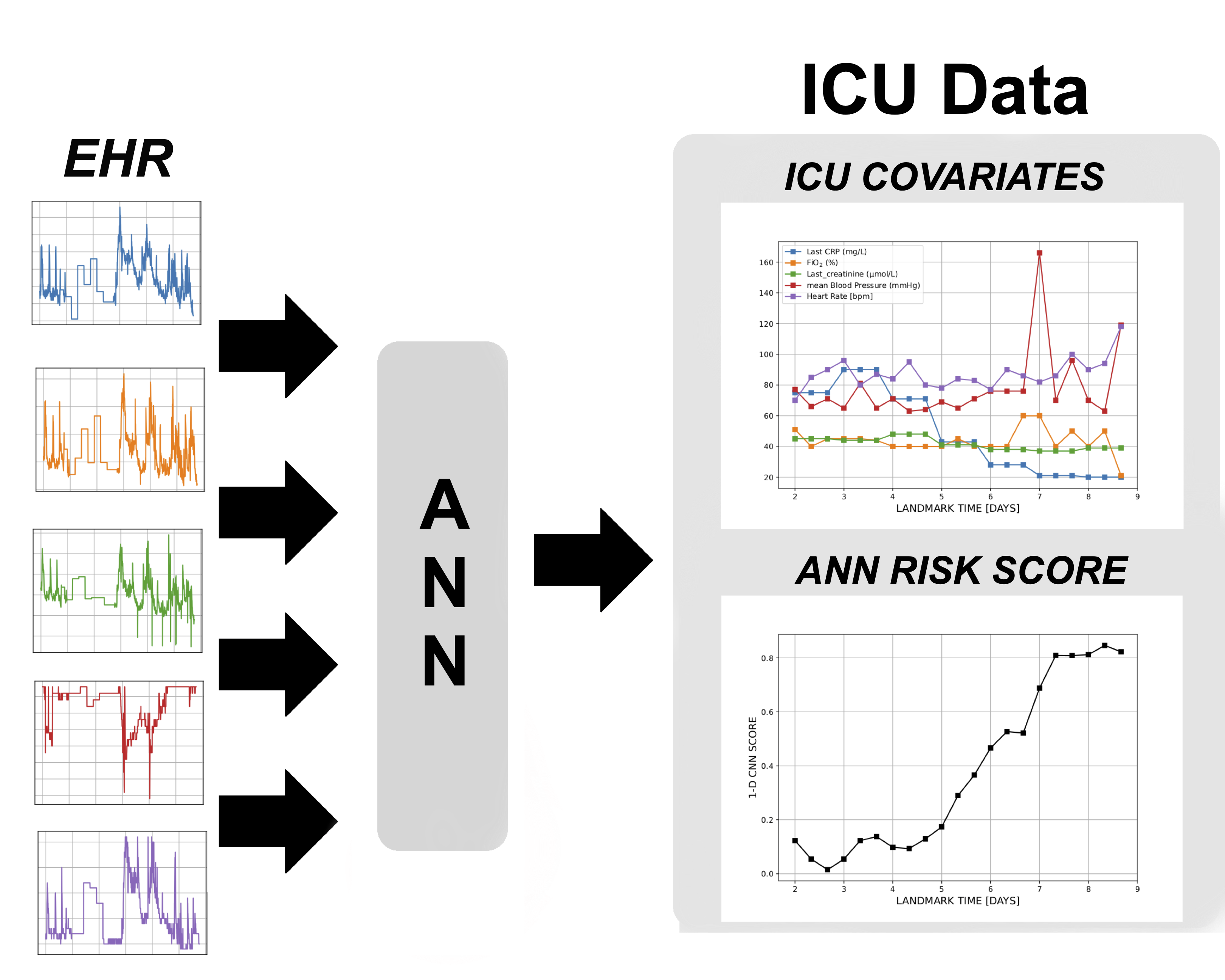}
    \caption{Schematic representation of the inclusion of the CNN-based risk score $Z_{\textnormal{CNN}}(t_{LM})$ in the ICU cohort data.}
    \label{fig:scheme_CNN_risk_score_to_ICU_data}
\end{figure}
A scheme of how we incorporated the risk score into the ICU predictors is illustrated in Figure~\ref{fig:scheme_CNN_risk_score_to_ICU_data}: for a single patient, the score is calculated for each LM time $t_{LM}$. At each $t_{LM}$ the values of other time-dependent covariates are reported as well (e.g., CRP, $\textnormal{Fi}\textnormal{O}_2$, creatinine level, mean blood pressure, mean heart rate).


\section{Step 2: \emph{Deep} LM-CR model}\label{icuai:landmark_model_J_causes}
\subsection{Notations and LM-CR model} \label{notations}
In this Section, we shall present the LM  model following the notation used in \cite{nicolaie2013dynamic}.

We consider a cohort consisting of $N$ subjects, and we denote with $\tilde{T}$ the time of failure, $C$ the censoring time, $D$ the cause of failure, and $\mathbf{Z}(\cdot)$ and an array of covariates.
In a general framework, a subject can only experience one of $J$ mutually excluding competing causes of failure; when it occurs $D$ takes a value in \{1, \dots, J\} corresponding to the cause under the exam.
Alternatively, when no cause has been experienced yet, $D$ always takes value $0$.
For the $i$-th subject, the tuple ($T_i, \Delta_i, \mathbf{Z}_{i}(\cdot)$) represents respectively the observed time $T_i = \min{(\tilde{T}_i, C_i)}$ (i.e.,  the earliest of failure and censoring time), the cause of failure $\Delta_i = \mathbf{1}(\tilde{T}_i < C_i)D_i$ (with $\mathbf{1}(\cdot)$ the indicator function), and $\mathbf{Z}_{i}(\cdot)$ the covariates up to time $T_i$.  
Note that $\Delta_{i} = 0$ denotes that the patient has experienced no failing causes; its clinical history has been censored.
We shall adopt the subscript $j$ to refer to the competing causes of failure,  with $j \in \{1, \dots, J\}$. 

We would like to derive a dynamic prediction of the probability distribution function of the failure time of cause $j$ at some time horizon ($t_{hor}$), conditional on surviving event-free and on the information available at a fixed time $t_{LM}$ (\emph{landmark time}).
More specifically, given a prediction window $w$ (such that $t_{hor}= t_{LM}+w$)
we would like to estimate the survival probability and the Cumulative
Incidence  Function (CIF) of cause $j$:
\begin{equation}
    S_{LM}(t_{hor}|\mathbf{Z}(t_{LM}), t_{LM}) := \mathbb{P}(T>t_{hor}| \mathbf{Z}(t_{LM}), t_{LM}),
\end{equation}
\begin{equation}
    F_{j, LM}(t_{hor}|\mathbf{Z}(t_{LM}), t_{LM}) := \mathbb{P}(T\leq t_{hor}, \Delta= j|  \mathbf{Z}(t_{LM}), t_{LM} ).
\label{eq:CIF}
\end{equation}

The LM approach consists of two steps:
\begin{enumerate}
    \item We first divide the time domain of our observations $[s_0, s_1]$ into  $n$ equi-spaced landmark points denoted with $\{t_{LM}^{k}\}_{k=1}^n$, where $t_{LM}^{1}\equiv s_0$ and $t_{LM}^{n}\equiv s_1$.
   We fix the width of the prediction window $w$ (i.e., the \emph{lead time}), and then for each LM time $t_{LM}^k$  we create a dataset by selecting all the subjects at risk at time $t_{LM}^k$ and by imposing  \emph{administrative} right-censoring at the time $t_{LM}^k+w$ (\emph{horizon time}). 
   Thus, for a vector of time-dependent covariates $\mathbf{Z}(t)$, only the values $\mathbf{Z}(t_{LM}^k)$ at $t_{LM}^k$ are considered in the $k$-th dataset. Finally, we create an extensive dataset by stacking all the datasets extracted at each landmark time $t_{LM}^k$ (LM \emph{super-dateset}). 

    \item The second step consist of fitting the \emph{LM-CR super-model} on the stacked LM \emph{super-dateset} \cite[]{nicolaie2013dynamic}.  
    Since at each $t_{LM}^k$, the vector $\mathbf{Z}(t_{LM}^k)$ is treated as a time constant vector of covariates, the dataset can be analyzed by using standard survival analysis methods. 
    \end{enumerate}

In the \emph{LM-CR super-model} we fit indeed a Cox proportional hazard model for the cause-specific hazard $\lambda_j$:
\begin{equation}
    \label{eq:lm_cause_specific_hazard_Cox}
    \lambda_{j}(t|t_{LM}, \mathbf{Z}(t_{LM})) = \lambda_{0j}(t|t_{LM})\exp{[\beta^{T}_{j}(t_{LM})\mathbf{Z}(t_{LM})]},
\end{equation}
where $\lambda_{0j}(t|t_{LM})$ denotes the (unspecified) baseline hazards and $\beta_{j}(t_{LM})$ the set of regressors specific for the $j$-th cause in within the interval interval $[t_{LM}, t_{LM}+w]$.
We assume that the coefficients $\beta$ depend on $t_{LM}$ in a smooth way, i.e., $ \beta_{j}(t_{LM}) = f_{j}(t_{LM}, {\beta}^{(0)}_j)$
with ${\beta}_j^{(0)}$ a vector of regression parameter  and $f_{\beta}(\cdot)$ a parametric function on time, e.g., a spline. Our choice has been a quadratic function:
 $$
 \beta_{j}(t_{LM}) := {\beta}_j^{(0)}+ \beta_{j}^{(1)}t_{LM}+  \beta_{j}^{(2)}t_{LM}^2.
$$
The estimation of $\lambda_{0j}(t|t_{LM})$ can be made through Breslow-type estimator; we can model such a dependence as
\begin{equation}\label{eq:Breslow-type_estimator}
    \lambda(t|t_{LM})_{0j}= \lambda_{0j}(t)\exp(\gamma_{j}(t_{LM})).
\end{equation}
As for the coefficients $\beta$, we assume the coefficients $\gamma$ of \eqref{eq:Breslow-type_estimator} to be parametrically dependent on the landmark times, e.g. by means of a quadratic spline
$$\gamma_{j}(t_{LM}) := {\gamma}_j^{(0)}+ \gamma_{j}^{(1)}t_{LM}+  \gamma_{j}^{(2)}t_{LM}^2.$$


Fitting this model with the Breslow partial likelihood for tied observations is equivalent to maximizing the pseudo-partial log-likelihood, as shown in \cite[]{nicolaie2013dynamic}. 
The landmark supermodel can be then fitted directly by applying a simple Cox model to the stacked data set.
Hence, after estimating the coefficients and the baseline cause-specific hazards, we get the \emph{plug-in} estimators for the survival probabilities (i.e., $\hat{S}_{LM}(t_{hor}|\boldsymbol{Z}(t_{LM}), t_{LM})$) and of the CIF of cause $j$ (i.e., $\hat{F}_{j, LM}(t_{hor} |\boldsymbol{Z}(t_{LM}), t_{LM})$).
The explicit form of these estimators is the following:
\begin{equation}\label{eq:lm_survival_final_estimation}
    \hat{S}_{LM}(t_{hor}|\boldsymbol{Z}(t_{LM}), t_{LM}) = \exp{ \left(-\sum_{j= 1}^{J}\exp{ (\boldsymbol{Z}(t_{LM})\hat{\beta}_{j}(t_{LM})+\hat{\gamma}_{j}(t_{LM}))} \left[\hat{\Lambda}_{0j}(t_{hor})- \hat{\Lambda}_{0j}(t_{LM})\right] \right)},
\end{equation}
and 
\begin{equation}\label{eq:lm_cum_incidence_function}
    \hat{F}_{j, LM}(t_{hor} |Z(t_{LM}), t_{LM}) = \sum_{t_{LM}< t_i \le t_{hor}} \hat{\lambda}_{0j}(t_i|\boldsymbol{Z}(t_{LM}))\hat{S}_{LM}(t_{hor}|\boldsymbol{Z}(t_{LM}), t_{LM}).
\end{equation}
The \emph{estimated cause-specific baseline} of  \eqref{eq:lm_cum_incidence_function} is given by
\begin{equation}
    \hat{\lambda}_{0j}(t_i) = \frac{\#(t_{LM}\le t_i \le t_{hor}, \Delta_i= j)}{\sum_{t_{LM}:t_{LM}\le t_i \le t_{hor}}\sum_{t_k:t_{LM}\le t_i \le t_k \le t_{hor}} \exp{[ \boldsymbol{Z}_{k}(t_{LM})^{T}\hat{\beta}_{j}(t_{LM})+\gamma_{j}(t_{LM})]}},
\end{equation}
while the \emph{estimated cause-specific cumulative baseline} is simply 
$$\hat{\Lambda}_{0j}(t) = \sum_{t_{i} \le t} \hat{\lambda}_{0j}(t_i).$$
\subsection{LM-CR for ICU-AI}\label{icuai:Deep-LCPH}
\begin{figure}
    \centering
    \includegraphics[width= 5cm]{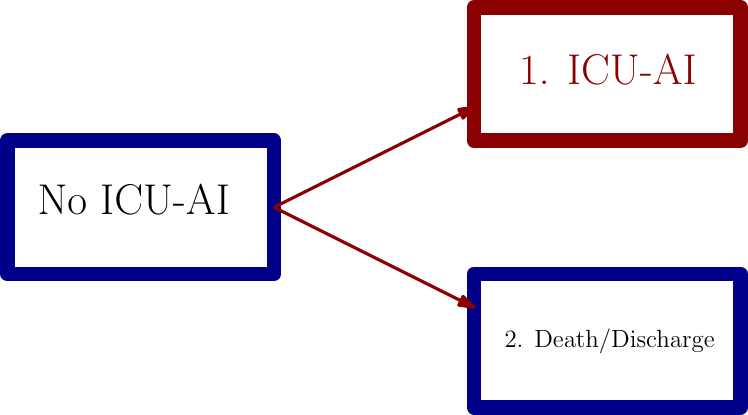}
    \caption{Competing risks model for ICU-AI.}  
        \label{fig:scheme_LCHP}
\end{figure}

In the context of dynamic predictions for  {ICU-AIs}, we adopted a CR model with three causes of failure:  \emph{ICU-AI}, \emph{death in the ICU}  and \emph{discharge}; see Figure~\ref{fig:scheme_LCHP}. No right censoring is present in the data, since no patient left the ICU before discharge or death.

Following the notation used in Section \ref{notations}, we denote with $\tilde{T}$ the time of failure, $D$ the cause of failure (i.e., $D= 1$ denotes an {ICU-AI}, while $D= 2$ discharge or death), and $\mathbf{Z}(\cdot)$ the array of covariates.
For the $i$-th subject the triple ($T_i, \Delta_i, \mathbf{Z}_{i}(\cdot)$) denotes the observed time $T_i \equiv \tilde{T}_i$, the cause of failure $\Delta_i \equiv D_i$, and $\mathbf{Z}_{i}(\cdot)$ the vector of covariates.

In this article, we consider the prediction window was set to $w= 24$ hours. The time domain is  $[s_0, s_1]$, with $s_0=48$ hours and $s_1=240$ hours, and we consider $n=25$ LM times $t_{LM}$, i.e., two subsequent LM times are at a distance of 8 hrs.

If we denote with $Z_{\textnormal{CNN}}(t)$  the CNN risk score at time $t$  (see Figure~\ref{fig:scheme_CNN_risk_score_to_ICU_data}) and with $\mathbf{Z}(t)$ the vector of all the other covariates in the LM-CR model at time $t$, we are interested at the dynamic predictions of the two models:
\begin {enumerate}
\item  \label{item:pred1}
$\pi_1:={F}_{1, LM}(t_{hor} |\mathbf{Z}(t_{LM}), t_{LM})$: i.e.,  the CIF of infection conditioned on the survival up to time $t_{LM}$ and on the \emph{low frequency} covariates (LM-CR model);
\item  
$\pi_2:={F}_{1, LM}(t_{hor} |\mathbf{Z}(t_{LM}), Z_{\textnormal{CNN}}(t_{LM}), t_{LM})$: the CIF of infection conditioned on the survival up to time $t_{LM}$ on both the \emph{low frequency} covariates and $Z_{\textnormal{CNN}}$ ({Deep}-LM-CR model).
\end{enumerate}

By comparing the accuracies of $\pi_1$ and $\pi_2$, we can measure the added predictive power of the CNN score. 
We shall refer to the first model with LM-CR and the second with Deep-LM-CR.

\subsection{Evaluation of LM-CR model}\label{sec:evaluation}

We use the {AUROC} metric to evaluate the prediction made at each single landmark time. 
When considering an overall measure, the evaluation of a global {AUROC} needs to consider the time-dependent character of the dynamic.
Similarly to the estimator of the prediction error proposed in \cite{spitoni2018prediction}, the evaluation of the overall  {AUROC} needs to take into account 
 the change in time of the size of the risk-set.
The absence of censoring allows us to estimate the overall {AUROC} score simply by:
\begin{equation}
    \operatorname{AUROC_{global}}  = \frac{\sum_{k = 1}^{n} R(t_{LM}^k)\operatorname{AUROC}(t_{LM}^k)}{\sum_{k= 1}^{n} R(t_{LM}^k)},
\end{equation}
with $t_{LM}^k$ the $k$-th landmark time, $n$ the total number of landmark times, and $R(t_{LM}^k)$ the size of the risk-set at time  $t_{LM}^k$.
Likewise, we estimated the overall Brier score as:
\begin{equation}
    \operatorname{BS_{global}}  = \frac{\sum_{k = 1}^{n} R(t_{LM}^k)\operatorname{BS}(t_{LM}^k)}{\sum_{l= 1}^{n} R(t_{LM}^l)}.
\end{equation}

The influence of the individual predictor in the prediction has been visualized by means of heat maps.
We compute the relative variation of the overall {AUROC} between the model including all predictors and the one where the predictor is removed.
Thus, we construct a heat map representing the relative change in {AUROC} due to the removal of a single predictor at landmarking time $t_{LM}$.

Finally, we remark that internal validation was performed using a \emph{10-fold cross-validation} method.
The  overall $\operatorname{AUROC_{global}}$ and the $\operatorname{AUROC}(t_{LM}^k)$, evaluated at each time $t_{LM}^k$, are averaged over the 10 folds.
In both the CR-LM model and the Deep-CR-LM model, we report 95\% bootstrap confidence intervals.

\subsection{Results}\label{icuai:dynamic_pred_icumai}
In this Section we shall show how the CNN risk score $Z_{\textnormal{CNN}}$ adds extra predictive information to the model, not present in the standard covariates. 

\begin{figure} 
    \centering
    \includegraphics[width= \textwidth]{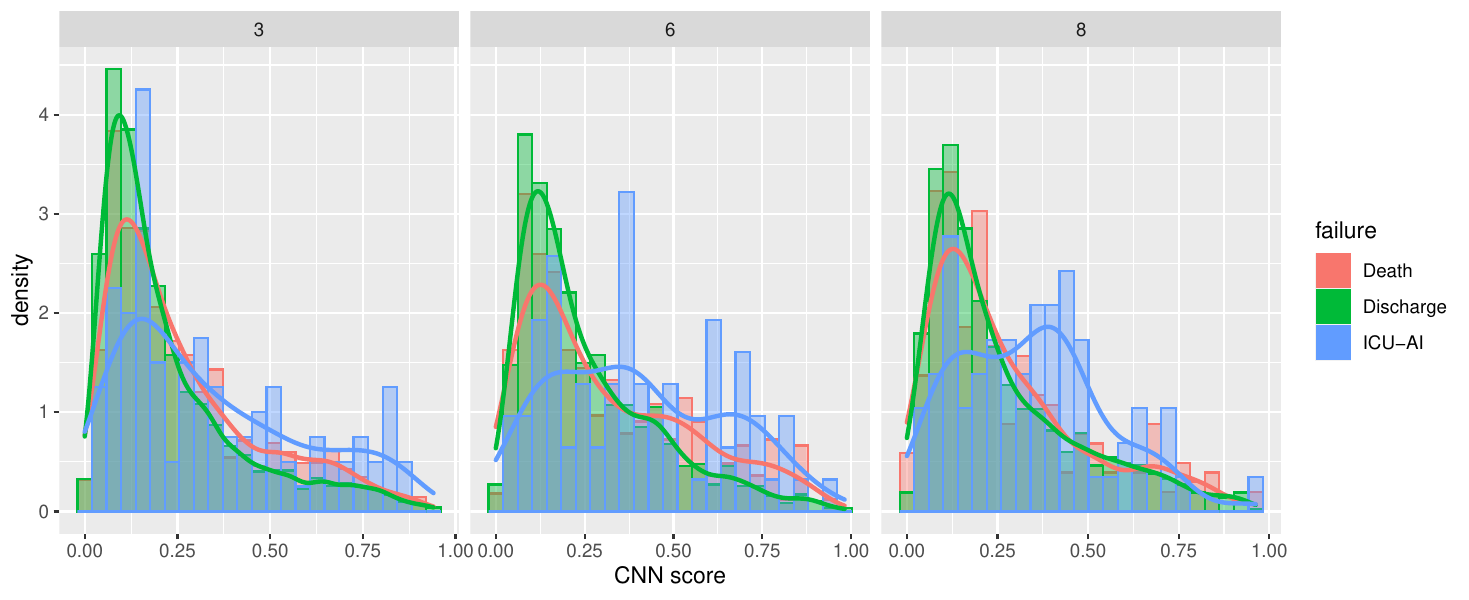}
    \caption{Distribution of the CNN risk score at three different landmark points ($t_{LM}^k\in\{ 3,6,8\}$ days), stratified for the cause of failure.}
    \label{fig:scores}
\end{figure}

In Figure~\ref{fig:scores} 
we plotted the empirical distribution of $Z_{\textnormal{CNN}}(t_{LM})$ for three landmark points  (i.e., $t_{LM}\in \{3,6,8\}$) and stratified by the cause of failure. 
As expected, the distribution of  $Z_{\textnormal{CNN}}$ for infected patients is more skewed on the right: while on day three this phenomenon is mild, on days 6 and 8 the skewness of the density distribution is much more evident.

\begin{figure} 
    \centering
    \includegraphics[width= \textwidth]{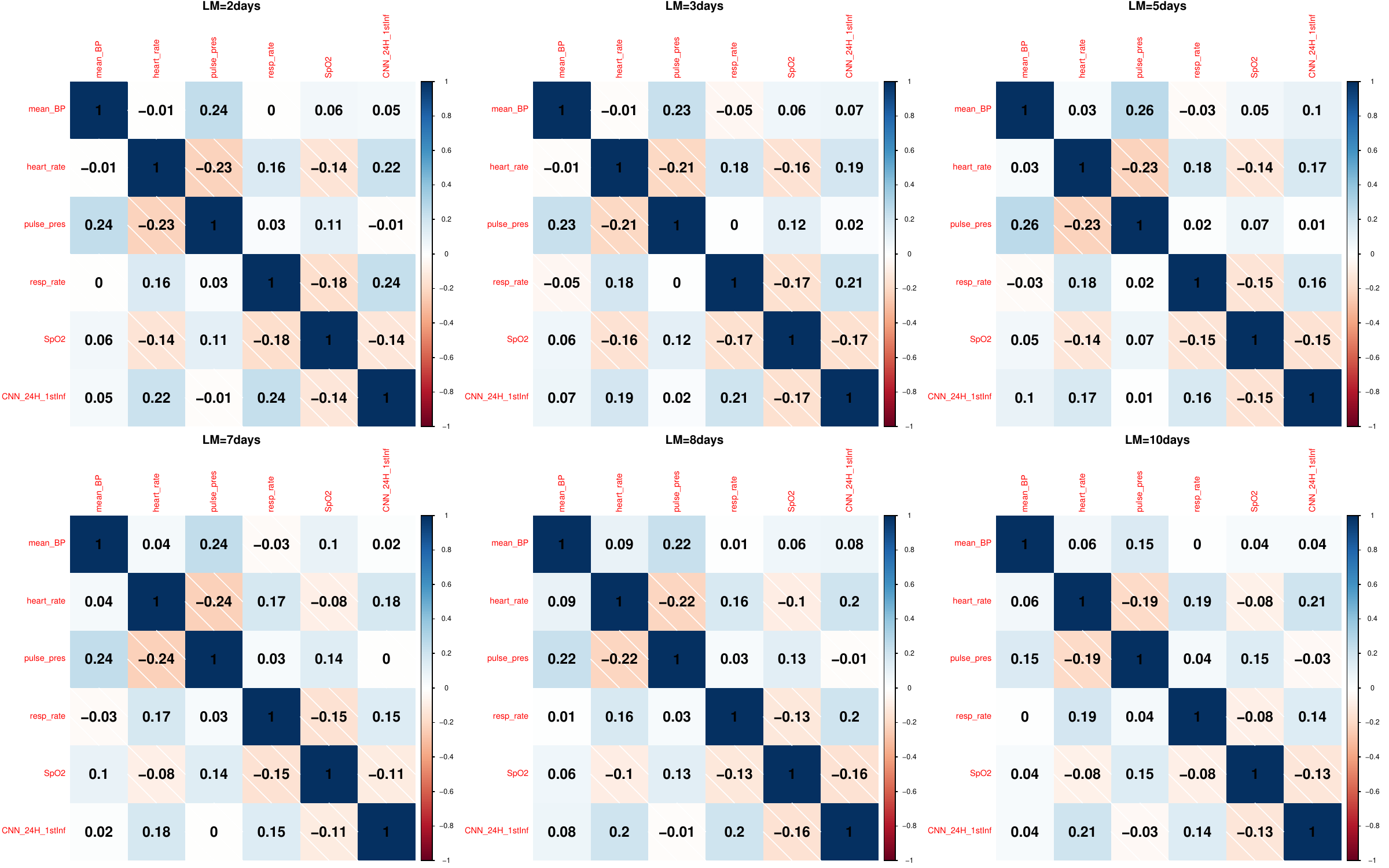}
    \caption{Correlation plot: CNN risk score vs. the vital signals (averaged in the 24 hours before the landmark). }
    \label{fig:corre}
\end{figure}

In Figure~\ref{fig:corre}, we reported the Pearson correlations between the CNN risk score and the vital signals averaged per 24-hour time windows prior to the landmark time
(i.e., the time-dependent covariates included in the LM-CR). 
Although the risk score is evaluated relative to these signals, only mild correlations are present. 
Our main hypothesis is indeed that $Z_{CNN}(t_{LM})$ has added predictive information, not contained in the other covariates $\mathbf{Z}(t_{LM})$.

Moreover, with regards to the cause-specific hazards for infection, the CNN risk score turned out to be the most important predictor: $\beta^{(0)}_{1; \textnormal{CNN}}=4.8$ (95\%CI 3.05-6.72). 
A complete list with all cause-specific hazards for ICU-AI is reported in Table~3 of the \emph{Supplementary Material}.

\begin{figure}
    \centering
    \includegraphics[width= 13cm]{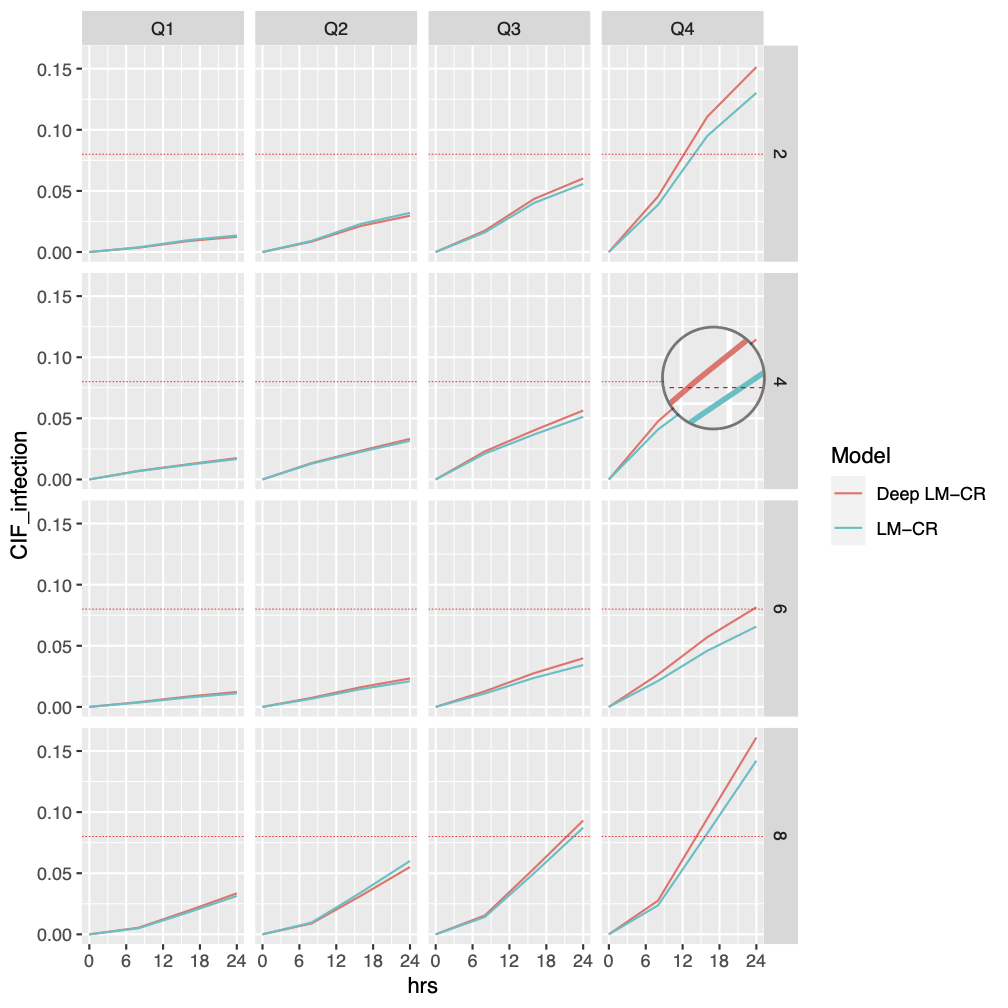}
    \caption{Comparison of the CIFs at different landmark times (i.e., $t_{LM}^k\in \{ 2,4,6,8\}$ days) of the models LM-CR and Deep-LM-CR.}
    \label{fig:lead}
\end{figure}

The LM approach provides a \emph{plug-in} estimator for the dynamic prediction (\ref{eq:CIF}) of the CIFs of ICU-AI. 
To give an example of the dynamic prediction allowed by the model, we reported in Figure~\ref{fig:lead} the CIFs for the LM-CR and the Deep-LM-CR models as a function of both the landmark time and the quantile  groups of the fitted linear predictors. Given the value of the covariates at the landmark time $t_{LM}$, the CIF at any $s$, with $s\in[t_{TM},t_{LM}+w]$ is given indeed by the \emph{plug-in} estimator $\hat{F}_{1, LM}(s |\boldsymbol{Z}(t_{LM}), t_{LM})$ of (\ref{eq:CIF}).

The dashed red line in Figure~\ref{fig:lead}  denotes an arbitrary warning level for the CIF of infection (e.g., $8\%$). We can see that, for the fourth quantile $Q_4$ and at LM time $t_{LM}=4$ days, the Deep-LM-CR model has a \emph{lead time} of circa 3 hours in reaching the warning threshold before the LM-CR model.

\begin{figure}
    \centering
    \includegraphics[width= 11cm]{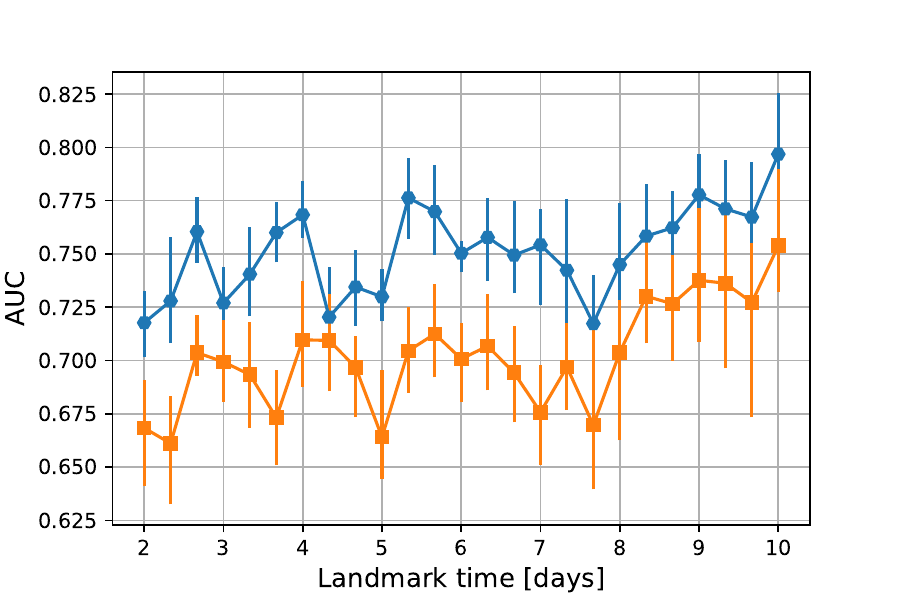}
    \caption{AUROC score (y-axis) as a function of the landmark times(x-axis). 
    The two curves represent the predictive performance of the basic  CR-LM model (orange) and the Deep-CR-LM model (blue). The error bars denote the 95\% bootstrap confidence intervals.}
    \label{fig:ICUFAI_AUROC}
\end{figure}
\begin{figure}
    \centering
    \includegraphics[width= 9cm]{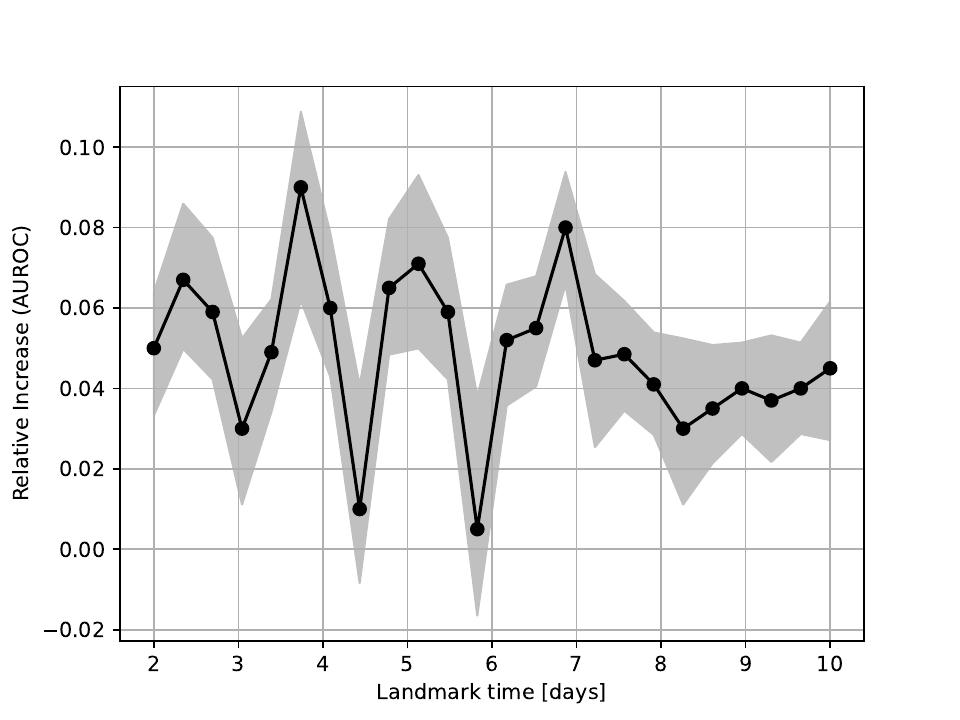}
    \caption{Overall relative increase of {AUROC} score (y-axis) as a function of the landmark times (x-axis) when including CNN-based risk score. The 95\% CI is represented as the light gray area.}
    \label{fig:ICUFAI_Rel_AUROC}
\end{figure}
\begin{figure}
    \centering
    \includegraphics[width= 11cm]{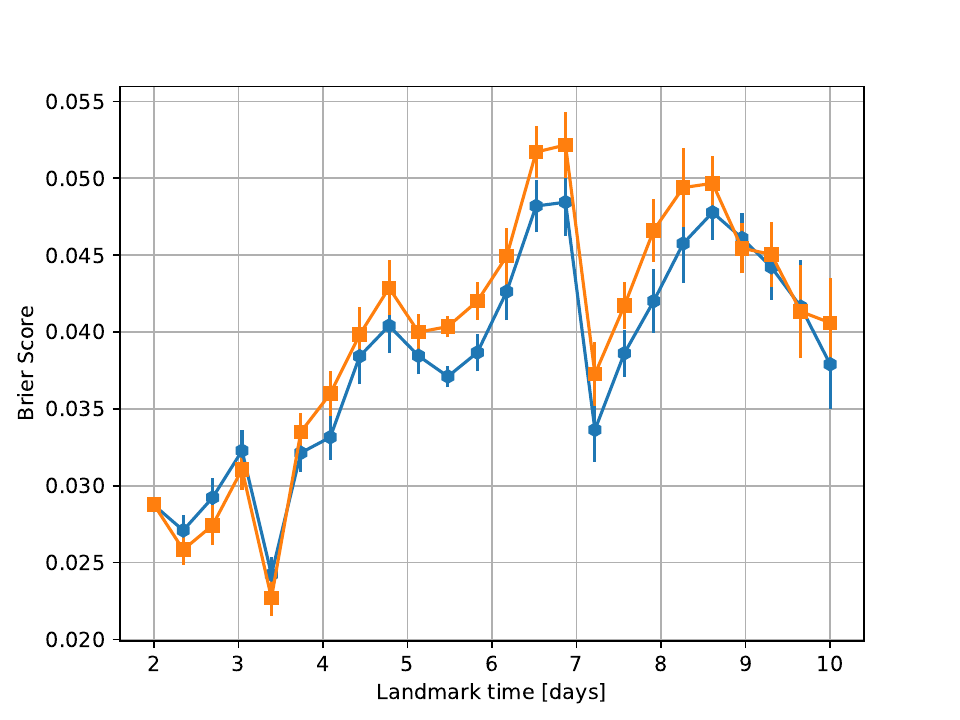}
    \caption{Brier score (y-axis) as a function of the landmark times (x-axis).
    In orange, the predictive power of the CR-LM model is shown, while the Deep-CR-LM is in blue. The error bars represent the 95\% bootstrap confidence interval.}
    \label{fig:Brier_Score_models}
\end{figure}
\begin{figure}
    \centering
    \includegraphics[width= 11cm]{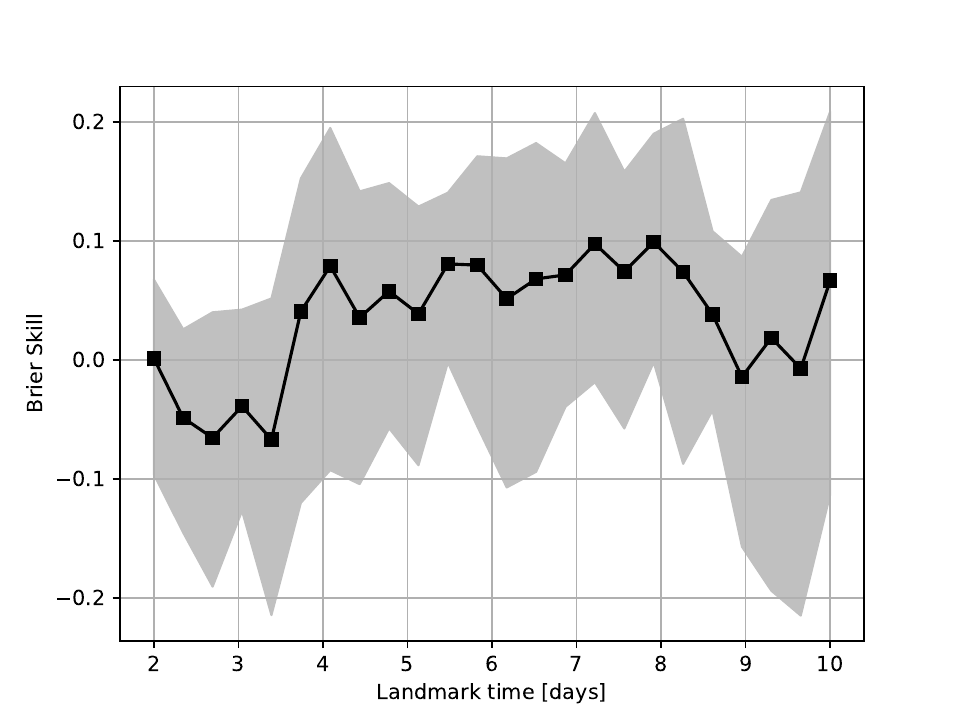}
    \caption{Brier skill (y-axis) as a function of the landmark times (x-axis).
    The 95\% CI is represented as the light grey area.}
    \label{fig:Brier_Skill_models}
\end{figure}

\begin{figure}
    \centering
    \includegraphics[width=\textwidth]{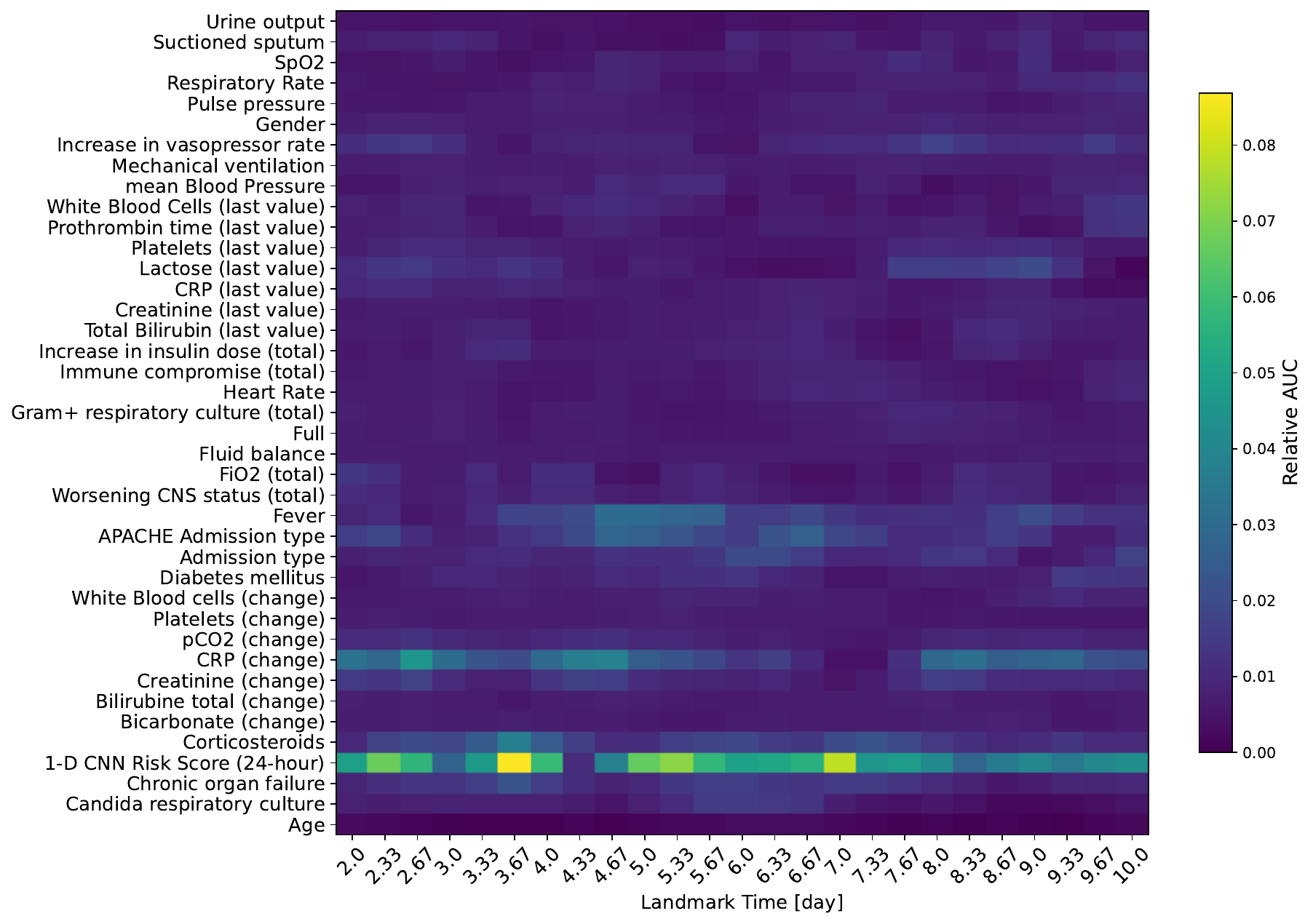}
    \caption{AUC heat-maps evaluating the impact of each predictor in the Deep-LM-CR model when predicting {ICU-AI}. The color of each pixel denotes the magnitude of the impact (relative {AUROC} increase) of one covariate (y-axis) for the LM time (x-axis) .}
    \label{fig:ICUFAI_heatmaps}
\end{figure}


The overall measure for the LM-CR model  is $\operatorname{AUROC_{global}}=0.69$ (95\%CI 0.68-0.70), while for the Deep-LM-CR  is $\operatorname{AUROC_{global}}=0.75$ (95\%CI 0.73-0.76).
The $\operatorname{AUROC}(t_{LM}^k)$ scores evaluated at each time $t_{LM}^k$, with $k\in\{1,\ldots,n\}$  are shown in Figure~\ref{fig:ICUFAI_AUROC}. The LM-CR model always shows lower predictive performance than the Deep-LM-CR. 
We noticed that at the early stage of the ICU stay (e.g., days 3.66) and around day 7, the {CNN} can improve the prediction of the traditional ICU clinical covariates of  about $8-9\%$, see Figure~\ref{fig:ICUFAI_Rel_AUROC}. 

The evaluation of the Brier Score revealed, for the LM-CR model, an overall measure $\operatorname{BS}_{global} = 0.037$ (95\% CI 0.036-0.039), while for the Deep-LM-CR we had $\operatorname{BS}_{global} = 0.036$ (95\% CI 0.035-0.037). 
The scores $\operatorname{BS}(t^{k}_{LM})$ of each landmark time, are shown in Figure \ref{fig:Brier_Score_models}.
For the majority of landmark times, we observed that the Deep-LM-CR was slightly more accurate than the other one; in contrast, the LM-CR turned out to be somewhat more precise in a few landmark times on days 2 and 4 and around days 9 and 10.
Such a result is also reported in Figure \ref{fig:Brier_Skill_models}; where the Brier Skill is shown. 
In the end, we observed an overall Brier Skill of 0.03 with 95\% CI equal to (0.01, 0.07).
The evaluation of all CI was accomplished via bootstrap resampling (bootstrap population equal to 1000 samples).

The impact of each explanatory variable $Z_j$ involved in the Deep-LM-CR model is shown in Figure~\ref{fig:ICUFAI_heatmaps}, in which we reported the heat-map of the relative increase in AUROC between the Deep-LM-CR without the covariate $Z_j$ and the full model (with $\mathbf{Z}(t_{LM})$ and $Z_{CNN}(t_{LM}))$. When $Z_j=Z_{CNN}$, we see that we observe a relative increase in AUROC of at least 
$4\%$.

In conclusion, an examination of the global performance of both the LM-CR and Deep-LM-CR models was also conducted, taking into account larger amplitude sliding windows of 16 and 24 hours.
In the case of the 16-hour sliding window, it was observed that the LM-CR exhibited an overall $\operatorname{AUROC_{global}}$ of 0.70 (95\% CI 0.69-0.71), whereas the Deep-LM-CR demonstrated a higher $\operatorname{AUROC_{global}}$ of 0.73 (95\% CI 0.72-0.74).
Similarly, for the 24-hour sliding window, the LM-CR displayed an overall $\operatorname{AUROC_{global}}$ of 0.69 (95\% CI 0.68-0.70), while the Deep-LM-CR exhibited a comparable $\operatorname{AUROC_{global}}$ of 0.73 (95\% CI 0.72-0.74).
A comparison between these findings and those obtained for models employing an 8-hour sliding window suggests a preference for the latter, as it attains the highest AUC when integrating the risk scores. 
It is noteworthy that the 8-hour sampling frequency stands as the maximum among low-frequency ICU covariates. 
Consequently, this analysis was constrained to datasets with sampling periods equal to or larger than 8 hours.

Summing up, we have shown that the two-step modeling can effectively lead to an increase in the accuracy of the predictions.  The extra predicting power comes from the inclusion of the CNN-based risk score, which is a summary measure of the predicting patterns found by the CNN model trained on only five vital signs signals (sample frequency of 1 minute). 

We remark that in our analysis we did not consider recurrent infections, but we limited the attention to the first episode of ICU-AI. 

\subsection{Comparison with a \emph{full} ANN-model}\label{icuai:TB-ANN}

For the sake of completeness, we would like to compare the predictions from our two-step modeling strategy with a full-ANN model. 
More specifically, we considered a Two-Branch Artificial Neural Network (TB-ANN) in order to process simultaneously high-frequency, low-frequency, and fixed-time covariates within a unique ANN model.
The TB-ANN consists of two distinct branches; the first has the scope of analyzing the high-frequency only, while the other analyses low-frequency and fixed covariates simultaneously.
The two branches are then connected and propagated through a prediction layer (i.e., a dense layer with a sigmoid activation function) returning an output score similar to the CNN risk score.
More details about the TB-ANN's architecture are discussed in the Supplementary Material.
Therefore, we are interested in the comparison between a model based on the estimation of interpretable quantities as hazard ratios (i.e., Deep LM-CR) and a completely ANN-based model (i.e., TB-ANN) for the prediction of impending infectious episodes.

\begin{figure}
    \centering\includegraphics[width= 11cm]{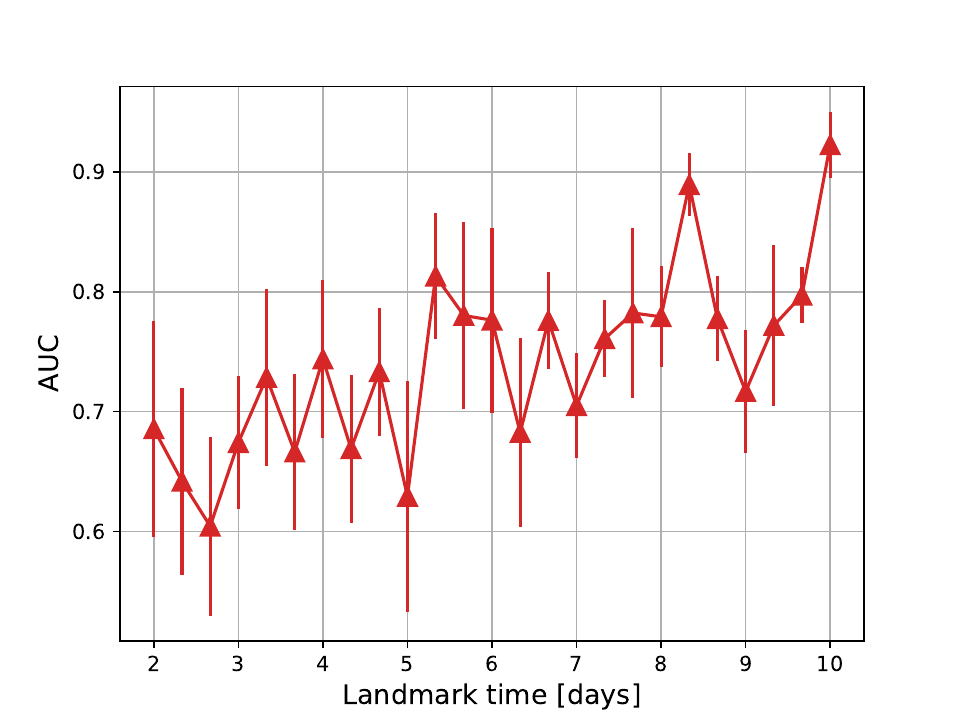}
    \caption{Mean value of AUROC of the TB-ANN at each landmark time. Error bars are the 95\% CI}
    \label{fig:AUROC_blackbox}
\end{figure}

\begin{figure}
    \centering\includegraphics[width= 11cm]{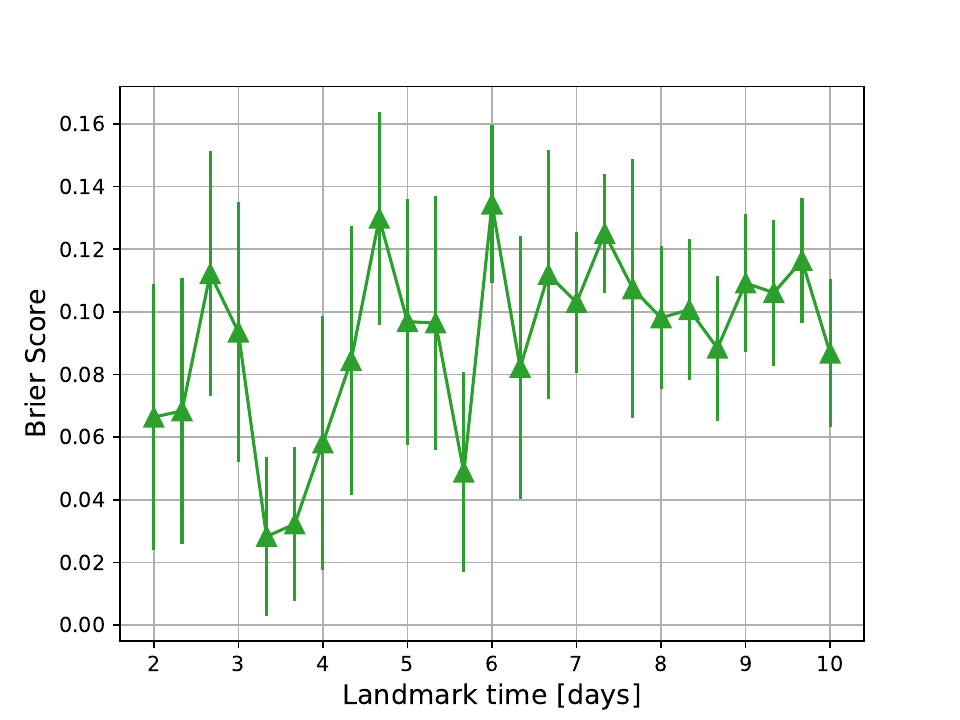}
    \caption{Mean value of Brier Score of the TB-ANN at each landmark time. Error bars are the 95\% CI}
    \label{fig:BS_blackbox}
\end{figure}

Similarly to the Deep LM-CR, we have trained the TB-ANN  at equi-spaced landmarking times within the time domain $[s_0, s_1]$, with $s_0$ and $s_1$ equal to 48 and 240, respectively; two generic subsequent landmarking times are 8-hour distant.
When training the TB-ANN models, we only considered the data available at each landmarking time to forecast the presence of an infectious episode in the next 24 hours.
To assess the TB-ANN predictive skill we primarily referred to the AUROC metric; we observed an overall AUROC (in the sense of Sec. \ref{sec:evaluation}) equal to 0.72 (95\% CI 0.55-0.9).
As a secondary metric, we also considered the Brier Score; we obtained an overall Brier Score of 0.08 (95\% CI 0.06-0.11).
The average values of AUROC and BS at each landmark time are reported, respectively, in figures \ref{fig:AUROC_blackbox} and \ref{fig:BS_blackbox}.
Error bars denote the 95\% confidence intervals

For most landmark times, we see that the TB-ANN's AUROC scores lay around values 0.7 and 0.8.
However, large fluctuations are present on days 2.67, 5, 8.33, and 10.
Especially for the last two mentioned, we have to remark that the reduction of the number of events at late landmark days might overestimate the AUROC scores. 
For the Brier Score, we observed different profiles at both early and late landmark days.
In fact, in the region 2-6 days the Brier Score presented important fluctuations around the global value of 0.08.
In particular, on days 3.33 and 3.66, we observed a score of 0.03, while higher scores larger than 0.10 were observed on days 2.66, and 4.66.
In contrast with this, the region 6-10 days appeared more stable, with much lower fluctuations, around the value of 0.10.  





The Deep LM-CR revealed a more stable prediction than the TB-ANN, while the accuracy was similar. 
However, our two-step strategy allows an immediate interpretation of the impact of each low-frequency covariate on the prediction and offers the possibility of using methods for interpreting the activity of the CNN, as explained in Section~\ref{icuai:XAI_icufai}.

\section{Explainability of CNN-based prediction of ICU-AI}\label{icuai:XAI_icufai}

In this section, we present our attempt to make interpretable the activity of the {CNN}.  As shown in Section~\ref{icuai:dynamic_pred_icumai}, the CNN-based risk score has added predicting power to the LM-CR model. However, for the moment, we do not have any information about the saliency of the vital signs selected by CNN during the training. This knowledge might be crucial for shedding some light on the relation between the activity of pattern recognition of the network and the medical conditions of a patient when an {ICU-AI} is approaching. 

To investigate which characteristics of the pattern selected by the CNN, we use the so-called \emph{Explainable Artificial Intelligence} (XAI),  namely a class of methods designed to understand the decisions and the predictions formulated by ANN techniques \cite[]{phillips2020four, vilone2021notions, castelvecchi2016can}. 
In the last decade, {XAI} has turned out a fundamental tool to make various ANN applications more reliable and transparent \cite[]{al2022novel, neubauer2022explainable, jimenez2020drug, dave2020explainable, lancia2022physics, lancia2023learning}.
The scope of {XAI} is to contrast indeed the widespread \emph{black box} attitude that many users have when applying {ANN} techniques.

\subsection{Explanability via SMOE scale}\label{icuai:ICUFAI_SMOE}
\label{sub:ex}
A saliency map is a map acting on the activated features in the hidden layers, generally used for showing which parts of the input are most important for the network's decisions.
The Saliency Map Order Equivalent scale (SMOE) used in the present paper is based on the algorithm developed by \cite{mundhenk2019efficient}:  an efficient and non-gradient method based on the statistical analysis of the activated feature maps. For a more detailed description of the SMOE scale, we refer the reader to Section~3 of the \emph{Supplementary Material}.

We would like to use the saliency maps for selecting, in the original 24-hour time series, the most relevant 8-hour patterns. 
We stress the fact that the selection of an 8-hour sliding window for identifying the most salient patterns was purposefully made to provide readers with a clear illustration of the method.
As shown in the last section, 8 hours turned out to be a suitable time scale to investigate the clinical dynamics of patients.
Such a choice therefore aims for maximum alignment with the results demonstrated for the Deep LR-CR model

\begin{figure}
    \centering
   \includegraphics[width=.8\textwidth]{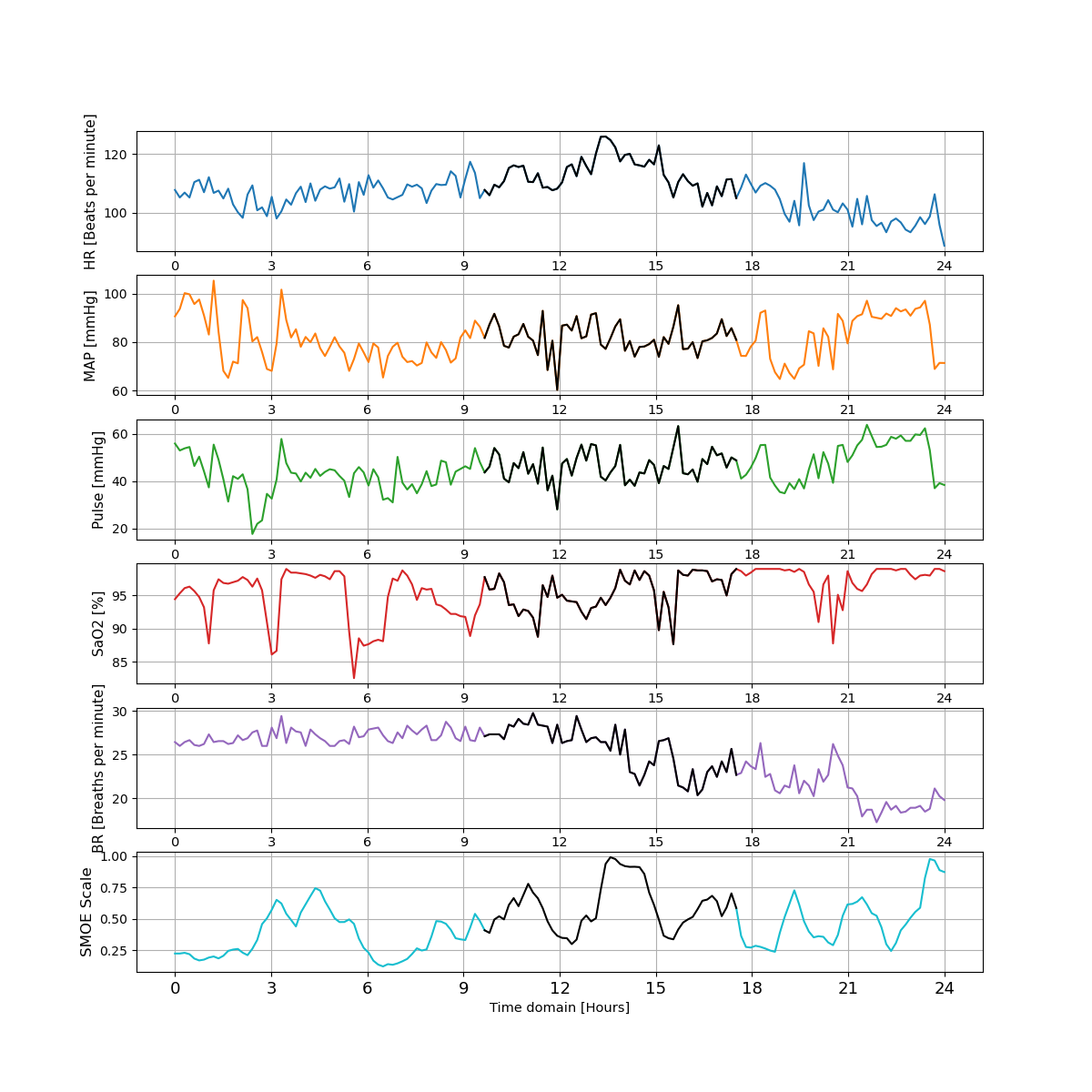}
    \caption{Schematic visualization of the 8-hour most salient patterns within a 24-hour time series sample.
    Starting from the top and descending, the first signals represent the vital signs considered.
    The cyan line represents the corresponding SMOE map.
    In black, the 8-hour chunk with the highest averaged SMOE value is outlined.}
    \label{fig:SMOE_example}
\end{figure}

The adopted approach is the following:
\begin{enumerate}

    \item We fit three different {CNNs}, one for each of  $t_{LM}^k\in\{3,7,10\}$. 
    We consider three distinct CNNs because the predicting patterns found by the network might differ among different periods of the ICU stay (see for instance the discussion in  Section~\ref{sub:res}). 
    The LM point \emph{3 days} is a proxy for an early time of the stay, \emph{7 days} for an intermediate time, and finally \emph{10 days} for a later moment. 
    The design of the networks is the same as described in Section~\ref{icuai:two_step_model_step1}. 
    All these models are validated via 5-fold cross-validation.
    
    \item We study the pattern recognition performed by the hidden layer, and we make it interpretable via the \emph{SMOE scale}.
    Through this method, we can visualize the regions of the input data with the highest saliency.
    Specifically, for each model developed at every LM time $t_{LM}^k$, we construct and visualize the saliency maps of the test set only.
    We repeat this action for each test set of each cross-validation fold. 
      
   
    \item  From each saliency map, we extract the \emph{8-hours interval} with the highest \emph{cumulative saliency value}. After having extracted the most relevant 8-hour patterns from each \emph{time series instance}, we can focus on their interpretation and their clustering. An example of the extraction of the 8-hour most salient pattern is shown in Figure~\ref{fig:SMOE_example}.
    \end{enumerate}

    %
    %
    %

\subsection{Data-driven clustering of salient patterns}
\label{sub:driven}
\begin{figure}
    \centering
    \includegraphics[width=1\textwidth]{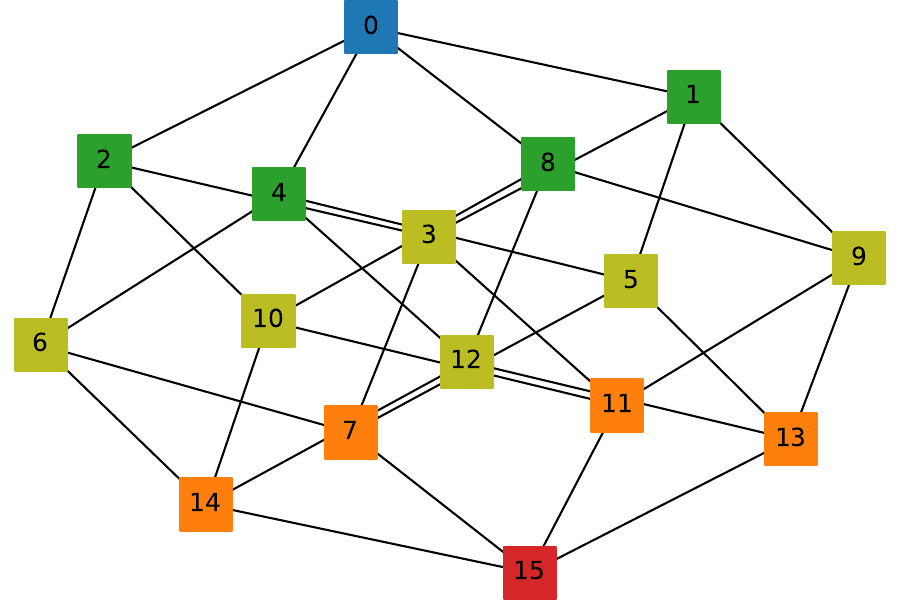}
    \caption{Illustration of the hypercube graph ($\mathcal{Q}_{4}$) with the 16 classes of the clustering. 
    The numbers on the nodes denote the classes as stated in Tab.\ref{tab:clusters}.
    The coloring of each node reflects the gravity of each clinical condition; (blue) No criticality, (green) one critical condition, (yellow) two critical conditions, (orange) three critical conditions, and (red) all four critical conditions}.
    \label{fig:graph_clusters}
\end{figure}

We focus now our attention on the clustering of the most salient patterns extracted in Section~\ref{sub:ex}. We would like indeed to answer the question: \emph{how can we link the activity of pattern recognition to some medical conditions, appearing when an {ICU-AI} is approaching? }
Our strategy for answering the question is the following:

\begin{enumerate}
    \item We collect the set of the most predictive patterns with an amplitude of 8 hours, obtained by applying the SMOE scale to the time series instances, as explained in Section~\ref{icuai:ICUFAI_SMOE}.
      \item We consider four clinical critical conditions, i.e., \emph{tachycardia}, \emph{hypotension}, \emph{desaturation}, and \emph{hyperventilation} (see Table \ref{tab:criteria}), which could predict the approaching of one {ICU-AI} episode. These medical conditions reflect the main symptoms of the Systemic Inflammatory Response Syndrome (SIRS), see \cite{chakraborty2019systemic}.
Tachycardia, hypotension, and hyperventilation are quite spread in the ICU, and they are usually mentioned in general guidelines for the ascertainment of {SIRS} \cite[]{comstedt2009systemic}. For the criteria reported in Table \ref{tab:criteria} we refer to \cite{comstedt2009systemic}; in specific for Desaturation, we refer to \cite[]{hafen2018oxygen}. 
    \item We evaluate the mean values of HR, ABP, $\operatorname{SaO_{2}}$ and BR for each  of the most salient  \emph{8-hour pattern} extracted via the SMOE scale.
    Depending on the values obtained (see the criteria in Table~\ref{tab:criteria}), we check the presence of the four clinical critical conditions.
    Thus, the combination of these conditions produces 16 different possible clinical situations of interest, as shown in Table~\ref{tab:clusters}: they represent the classes of the proposed data-driven clustering. In Figure~\ref{fig:graph_clusters} the 16 distinct classes are represented as nodes of a graph (i.e., a four-dimensional hypercube ($\mathcal{Q}_4$)).
    
    
\end{enumerate}

\begin{table}
    \centering
    \begin{tabular}{|c|c|}
    \hline
        \textbf{Critical Condition} & \textbf{Criterion} \\
        \hline
        Tachycardia      &  Hearth Rate $\ge$90 beats per minute \\
        \hline
        Hypotension      &  Arterial Blood Pressure (mean) $\le$ 80mmHg \\
        \hline
        Desaturation     &  $\operatorname{SaO_{2}}\le$ 95\%\\
        \hline
        Hyperventilation     &  Breath Rate $\ge$ 24 breaths per minute \\
        \hline
    \end{tabular}
    \caption{Critical conditions and their criteria.}
    \label{tab:criteria}
\end{table}

\begin{table}    
    \centering
    \begin{tabular}{|c|l|}
    \hline
        \textbf{Class} & \textbf{Data Driven Cluster (Clinical Conditions)} \\
        \hline
        0      &  None \\
        \hline
        1      &  Tachycardia \\
        \hline
        2      &  Hypotension \\
        \hline
        3      &  Hypotension, Tachycardia \\
        \hline
        4      &  Desaturation \\
        \hline
        5      &  Desaturation, Tachycardia \\
        \hline
        6      &  Desaturation, Hypotension\\
        \hline
        7      &  Desaturation, Hypotension, Tachycardia \\
        \hline
        8      &  Hyperventilation \\
        \hline
        9      &  Hyperventilation, Tachycardia \\
        \hline
        10      &  Hyperventilation, Hypotension \\
        \hline
        11      &  Hyperventilation, Hypotension, Tachycardia \\
        \hline
        12      &  Hyperventilation, Desaturation\\
        \hline
        13      &  Hyperventilation,  Desaturation, Tachycardia \\
        \hline
        14     &   Hyperventilation, Desaturation, Hypotension \\
        \hline
        15      &  Hyperventilation, Desaturation, Hypotension, Tachycardia\\
        \hline
    \end{tabular}
    \caption{List of the 16 clinical conditions (classes of the clustering).}
    \label{tab:clusters}
\end{table}

\begin{figure}
    \begin{subfigure}{0.50\textwidth}
    \includegraphics[width=0.80\textwidth]{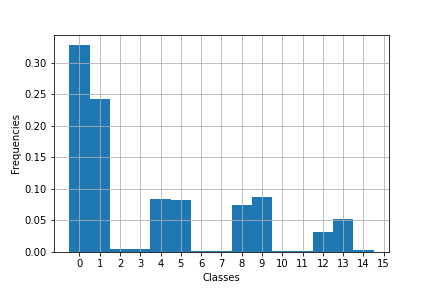}
    \caption{}
    \end{subfigure}
    \begin{subfigure}{0.50\textwidth}
    \includegraphics[width=0.80\textwidth]{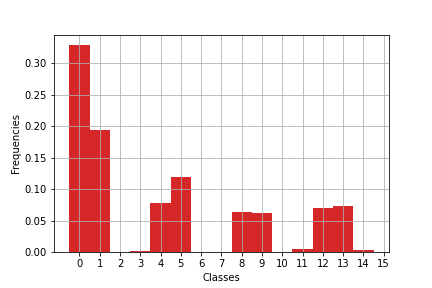}
    \caption{}
    \end{subfigure}
    \begin{subfigure}{0.50\textwidth}
    \includegraphics[width=0.80\textwidth]{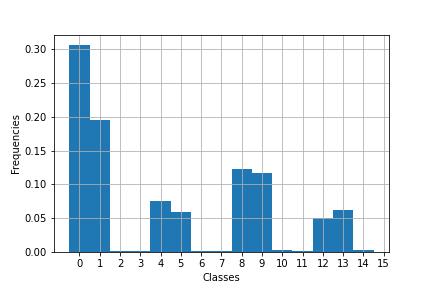}
    \caption{}
    \end{subfigure}
    \begin{subfigure}{0.50\textwidth}
    \includegraphics[width=0.80\textwidth]{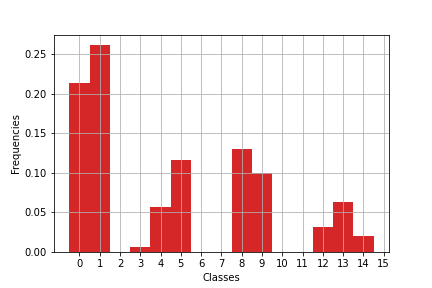}
    \caption{}
    \end{subfigure}
    \begin{subfigure}{0.50\textwidth}
    \includegraphics[width=0.80\textwidth]{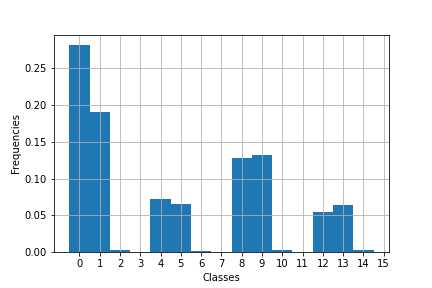}
    \caption{}
    \end{subfigure}
    \begin{subfigure}{0.50\textwidth}
    \includegraphics[width=0.80\textwidth]{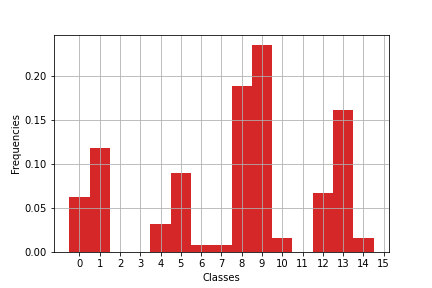}
    \caption{}
    \end{subfigure}
    \caption{Histograms the data-driven clustering approach. Bins on the x-axis represent the 16 classes. Blue histograms concern the non-infected instances, 
    whereas the red ones the infected instances. {CNN} trained on day 3 is described by (a) and (b), on day 7 by (c) and (d), and on day 10 by (e) and (f).}
    \label{fig:data_driven_cluster}
\end{figure}

\subsection{Results of the data-driven clustering}\label{sub:res}
Histograms with the relative frequencies of the 16 data-driven clusters are shown in Figure~\ref{fig:data_driven_cluster}.
For day 3 (see Figures~\ref{fig:data_driven_cluster}(a) and \ref{fig:data_driven_cluster}(b)),two-sample Kolmogorov-Smirnov test \cite[]{hodges1958significance}
reveals that the sample distributions of the classes between \emph{not-infected} and \emph{infected} instances are not significantly different (p-value=0.21).
However, we can observe a completely different scenario on both  days 7 and 10 (see Figure~\ref{fig:data_driven_cluster} (d)-(f)), where the null hypothesis of the two-samples Kolmorogov-Smirnov test is rejected (p-value= 0.0003 and p-value= $10^{-10}$  respectively). Hence, this  analysis shows that different clinical conditions could represent an essential feature of the patterns that the {CNN} model captures during the learning phase.
For instance, for \emph{infected} instances, at day 10, the prevalence of at least one of these 16 conditions is around 94\%, while 79\% at day 7; see Figure~\ref{fig:data_driven_cluster} (d)-(f)).
Precisely, on day 10, events with hyperventilation correspond at 70\% of samples, and in combination with tachycardia 23\%.
While a day 7 tachycardia is much more relevant and occurs in 50\% of infectious samples. Therefore, the most salient 8-hour subinterval of our \emph{time series instance} can be linked to precise medical conditions, which are known to be related to the presence of an ICU-AI. 

\section{Conclusions}

We have shown that the proposed two-step modeling of ICU-AI is simultaneously an accurate predicting tool and an interpretable model. As we have discussed, predicting an infection with our adopted definition is a challenging problem: the time to infection is determined by the start of an antibiotic treatment. Hence, the impossibility of determining the actual time of infection represents an intrinsic obstacle to building a performative prediction model based on high-frequency data.

However, the CNN can capture potentially predicting patterns by analyzing the time series of five vital sign signals. 
These patterns contain extra predictive information and they are only mildly correlated with the averaged quantities of the vital signals, routinely included in the traditional survival models.
Moreover, we have shown as well that the SMOE scale might help physicians in clustering patients with an approaching infection. 
%


In this work, we have considered a survival model without censoring, since ICU patients are fully monitored during their stay.
However, methods based on the pseudo-observations \cite[]{andersen2010pseudo} represent a solid strategy to contrast the biasing of the desired dynamic prediction due to the censoring data.
In the context of LM-based survival dynamic predictions,
such an approach has already been proposed; e.g., the work of  \cite{nicolaie2013dynamic}, and in a similar way \cite{cortese2013comparing}, presents a well-founded generalization of landmark models able to estimate how baseline and covariate effects lead to the desired dynamic predictions with left and right censoring.
Likewise, a first attempt to conjugate {ANN} and survival predictions have recently been proposed by \cite{zhao2020deep}.
Despite considering only a simple MLP architecture to solve a generalized model with a logit link, this work represents a promising approach for developing new methodologies for increasing the accuracy of the survival predictions obtained by a multiplex {ANN} architecture fed with censored data.
In comparison with the TB-ANN, we showed that an LM approach can lead to slightly more accurate and well-calibrated predictions.
Despite showing almost the same overall accuracy level, the TB-ANN predictions tend to be much more sensitive on different landmark days.
To our knowledge, this fact reflects the intrinsic difficulty of well-calibrating an {ANN} classifier when analyzing a vast amount of information coming from different data structures.

We have illustrated the methodology in a competing risks framework. 
However, the LM approach has recently been extended to \emph{multi-state} models, even without the Markov assumption \cite[]{MS_1,MS_2}. 
Therefore, as a further extension, we could model recurrent infections as new states in a non-Markov multi-state model, with transition hazards that might depend on the previous infections' sequence.
Moreover, another future challenging direction of investigation is a sort of  \emph{inversion} of the CNN, in order to identify and classify the patterns in the signal with higher predicting power. This analysis might help in performing a more precise clustering of the patients with fore-coming ICU-AI. 

\section*{Code Availability}
Python codes and modules are available on GitHub: \url{ https://github.com/glancia93/ICUAI-dynamic-prediction/blob/main/ICUAI_module.py}.

\section*{Declaration of Interest Statement}
The authors declare to have no conflicts to disclose.

\section*{Declaration of generative AI in scientific writing}

The authors declare that no generative AI and AI-assisted technologies have been utilized during the writing process of this manuscript.

\bibliographystyle{apalike}


\begin{thebibliography}{}

\bibitem[Al-Najjar et~al., 2022]{al2022novel}
Al-Najjar, H.~A., Pradhan, B., Beydoun, G., Sarkar, R., Park, H.-J., and
  Alamri, A. (2022).
\newblock A novel method using explainable artificial intelligence (xai)-based
  shapley additive explanations for spatial landslide prediction using
  time-series sar dataset.
\newblock {\em Gondwana Research}.

\bibitem[Andersen et~al., 1993]{bible}
Andersen, P.~K., Borgan, O., Gill, R.~D., and Keiding, N. (1993).
\newblock {\em Statistical Models Based on Counting Processes}.
\newblock Springer.

\bibitem[Andersen and Pohar~Perme, 2010]{andersen2010pseudo}
Andersen, P.~K. and Pohar~Perme, M. (2010).
\newblock Pseudo-observations in survival analysis.
\newblock {\em Statistical methods in medical research}, 19(1):71--99.

\bibitem[Borovykh et~al., 2017]{borovykh2017conditional}
Borovykh, A., Bohte, S., and Oosterlee, K. (2017).
\newblock Conditional time series forecasting with convolutional neural
  networks.
\newblock In {\em Lecture Notes in Computer Science/Lecture Notes in Artificial
  Intelligence}, pages 729--730.

\bibitem[Brier, 1950]{brier1950verification}
Brier, G.~W. (1950).
\newblock Verification of forecasts expressed in terms of probability.
\newblock {\em Monthly weather review}, 78(1):1--3.

\bibitem[Castelvecchi, 2016]{castelvecchi2016can}
Castelvecchi, D. (2016).
\newblock Can we open the black box of ai?
\newblock {\em Nature News}, 538(7623):20.

\bibitem[Chakraborty and Burns, 2019]{chakraborty2019systemic}
Chakraborty, R.~K. and Burns, B. (2019).
\newblock Systemic inflammatory response syndrome.

\bibitem[Comstedt et~al., 2009]{comstedt2009systemic}
Comstedt, P., Storgaard, M., and Lassen, A.~T. (2009).
\newblock The systemic inflammatory response syndrome (sirs) in acutely
  hospitalised medical patients: a cohort study.
\newblock {\em Scandinavian journal of trauma, resuscitation and emergency
  medicine}, 17(1):1--6.

\bibitem[Cortese and Andersen, 2010]{cortese2010competing}
Cortese, G. and Andersen, P.~K. (2010).
\newblock Competing risks and time-dependent covariates.
\newblock {\em Biometrical Journal}, 52(1):138--158.

\bibitem[Cortese et~al., 2013]{cortese2013comparing}
Cortese, G., Gerds, T.~A., and Andersen, P.~K. (2013).
\newblock Comparing predictions among competing risks models with
  time-dependent covariates.
\newblock {\em Statistics in medicine}, 32(18):3089--3101.

\bibitem[Dantes and Epstein, 2018]{dantes2018combatting}
Dantes, R.~B. and Epstein, L. (2018).
\newblock Combatting sepsis: a public health perspective.
\newblock {\em Clinical infectious diseases}, 67(8):1300--1302.

\bibitem[Dave et~al., 2020]{dave2020explainable}
Dave, D., Naik, H., Singhal, S., and Patel, P. (2020).
\newblock Explainable ai meets healthcare: A study on heart disease dataset.
\newblock {\em arXiv preprint arXiv:2011.03195}.

\bibitem[Deng et~al., 2023]{deng2023dynamic}
Deng, Y., Ma, Y., Fu, J., Wang, X., Yu, C., Lv, J., Man, S., Wang, B., and Li,
  L. (2023).
\newblock A dynamic machine learning model for prediction of nafld in a health
  checkup population: A longitudinal study.
\newblock {\em Heliyon}, 9(8).

\bibitem[Ferrer et~al., 2019]{miss}
Ferrer, L., Putter, H., and Proust-Lima, C. (2019).
\newblock Individual dynamic predictions using landmarking and joint modeling:
  Validation of estimators and robustness assessment.
\newblock {\em Statistical Methods in Medical Research}, 28(12):3649--3666.

\bibitem[Gandin et~al., 2021]{gandin2021interpretability}
Gandin, I., Scagnetto, A., Romani, S., and Barbati, G. (2021).
\newblock Interpretability of time-series deep learning models: A study in
  cardiovascular patients admitted to intensive care unit.
\newblock {\em Journal of Biomedical Informatics}, 121:103876.

\bibitem[Guo-yan et~al., 2019]{guo2019combined}
Guo-yan, X., Jin, Z., Cun-you, S., Wen-bin, H., and Fan, L. (2019).
\newblock Combined hydrological time series forecasting model based on cnn and
  mc.
\newblock {\em Computer and Modernization}, (11):23.

\bibitem[Hafen and Sharma, 2022]{hafen2018oxygen}
Hafen, B.~B. and Sharma, S. (2022).
\newblock {\em Oxygen saturation}.
\newblock StatPearls, StatPearls Publishing.

\bibitem[Hodges, 1958]{hodges1958significance}
Hodges, J.~L. (1958).
\newblock The significance probability of the smirnov two-sample test.
\newblock {\em Arkiv f{\"o}r Matematik}, 3(5):469--486.

\bibitem[Hoff et~al., 2019]{MS_2}
Hoff, R., Putter, H., Mehlum, I.~S., and Gran, J.~M. (2019).
\newblock Landmark estimation of transition probabilities in non-markov
  multi-state models with covariates.
\newblock {\em Lifetime Data Analysis}, 25(4):660--680.

\bibitem[Ivanov et~al., 2022]{ivanov2022accurate}
Ivanov, O., Molander, K., Dunne, R., Liu, S., Masek, K., Lewis, E., Wolf, L.,
  Travers, D., Brecher, D., Delaney, D., et~al. (2022).
\newblock Accurate detection of sepsis at ed triage using machine learning with
  clinical natural language processing.
\newblock {\em arXiv preprint arXiv:2204.07657}.

\bibitem[Jim{\'e}nez-Luna et~al., 2020]{jimenez2020drug}
Jim{\'e}nez-Luna, J., Grisoni, F., and Schneider, G. (2020).
\newblock Drug discovery with explainable artificial intelligence.
\newblock {\em Nature Machine Intelligence}, 2(10):573--584.

\bibitem[Kagaya et~al., 2014]{kagaya2014food}
Kagaya, H., Aizawa, K., and Ogawa, M. (2014).
\newblock Food detection and recognition using convolutional neural network.
\newblock In {\em Proceedings of the 22nd ACM international conference on
  Multimedia}, pages 1085--1088.

\bibitem[Kingma and Ba, 2015]{kingma2014adam}
Kingma, D.~P. and Ba, J. (2015).
\newblock Adam: A method for stochastic optimization.
\newblock In {\em ICLR}.

\bibitem[Klouwenberg et~al., 2013]{klouwenberg2013interobserver}
Klouwenberg, P. M.~K., Ong, D.~S., Bos, L.~D., de~Beer, F.~M., van Hooijdonk,
  R.~T., Huson, M.~A., Straat, M., van Vught, L.~A., Wieske, L., Horn, J.,
  et~al. (2013).
\newblock Interobserver agreement of centers for disease control and prevention
  criteria for classifying infections in critically ill patients.
\newblock {\em Critical care medicine}, 41(10):2373--2378.

\bibitem[Kwon et~al., 2018]{kwon2018empirical}
Kwon, D., Natarajan, K., Suh, S.~C., Kim, H., and Kim, J. (2018).
\newblock An empirical study on network anomaly detection using convolutional
  neural networks.
\newblock In {\em ICDCS}, pages 1595--1598.

\bibitem[Lancia et~al., 2023]{lancia2023learning}
Lancia, G., Durastanti, C., Spitoni, C., De~Benedictis, I., Sciortino, A.,
  Cirillo, E.~N., Ledda, M., Lisi, A., Convertino, A., and Mussi, V. (2023).
\newblock Learning models for classifying raman spectra of genomic dna from
  tumor subtypes.
\newblock {\em arXiv preprint arXiv:2302.08918}.

\bibitem[Lancia et~al., 2022]{lancia2022physics}
Lancia, G., Goede, I., Spitoni, C., and Dijkstra, H. (2022).
\newblock Physics captured by data-based methods in el ni{\~n}o prediction.
\newblock {\em Chaos: An Interdisciplinary Journal of Nonlinear Science},
  32(10).

\bibitem[Liu, 2018]{liu2018feature}
Liu, Y.~H. (2018).
\newblock Feature extraction and image recognition with convolutional neural
  networks.
\newblock In {\em Journal of Physics: Conference Series}, volume 1087, page
  062032. IOP Publishing.

\bibitem[Livieris et~al., 2020]{livieris2020cnn}
Livieris, I.~E., Pintelas, E., and Pintelas, P. (2020).
\newblock A cnn--lstm model for gold price time-series forecasting.
\newblock {\em Neural computing and applications}, 32(23):17351--17360.

\bibitem[Lou and Shi, 2020]{lou2020face}
Lou, G. and Shi, H. (2020).
\newblock Face image recognition based on convolutional neural network.
\newblock {\em China Communications}, 17(2):117--124.

\bibitem[Maki et~al., 2008]{maki2008nosocomial}
Maki, D.~G., Crnich, C.~J., and Safdar, N. (2008).
\newblock Nosocomial infection in the intensive care unit.
\newblock {\em Critical care medicine}, page 1003.

\bibitem[May et~al., 2011]{May11}
May, R., Dandy, G., and Maier, H. (2011).
\newblock Review of input variable selection methods for artificial neural
  networks.
\newblock In Suzuki, K., editor, {\em Artificial Neural Networks}, chapter~2.
  IntechOpen, Rijeka.

\bibitem[Mundhenk et~al., 2019]{mundhenk2019efficient}
Mundhenk, T.~N., Chen, B.~Y., and Friedland, G. (2019).
\newblock Efficient saliency maps for explainable ai.
\newblock {\em arXiv preprint arXiv:1911.11293}.

\bibitem[Naseer et~al., 2018]{naseer2018enhanced}
Naseer, S., Saleem, Y., Khalid, S., Bashir, M.~K., Han, J., Iqbal, M.~M., and
  Han, K. (2018).
\newblock Enhanced network anomaly detection based on deep neural networks.
\newblock {\em IEEE access}, 6:48231--48246.

\bibitem[Neubauer and Roy, 2022]{neubauer2022explainable}
Neubauer, M.~S. and Roy, A. (2022).
\newblock Explainable ai for high energy physics.
\newblock {\em arXiv preprint arXiv:2206.06632}.

\bibitem[Nicolaie et~al., 2013]{nicolaie2013dynamic}
Nicolaie, M., Van~Houwelingen, J., De~Witte, T., and Putter, H. (2013).
\newblock Dynamic prediction by landmarking in competing risks.
\newblock {\em Statistics in medicine}, 32(12):2031--2047.

\bibitem[Phillips et~al., 2020]{phillips2020four}
Phillips, P.~J., Hahn, C.~A., Fontana, P.~C., Broniatowski, D.~A., and
  Przybocki, M.~A. (2020).
\newblock Four principles of explainable artificial intelligence.
\newblock {\em Gaithersburg, Maryland}.

\bibitem[Proust-Lima and Taylor, 2009]{proust2009development}
Proust-Lima, C. and Taylor, J.~M. (2009).
\newblock Development and validation of a dynamic prognostic tool for prostate
  cancer recurrence using repeated measures of posttreatment psa: a joint
  modeling approach.
\newblock {\em Biostatistics}, 10(3):535--549.

\bibitem[Putter and Spitoni, 2018]{MS_1}
Putter, H. and Spitoni, C. (2018).
\newblock Non-parametric estimation of transition probabilities in non-markov
  multi-state models: The landmark aalen–johansen estimator.
\newblock {\em Statistical Methods in Medical Research}, 27(7):2081--2092.

\bibitem[Rizopoulos, 2011]{rizopoulos2011dynamic}
Rizopoulos, D. (2011).
\newblock Dynamic predictions and prospective accuracy in joint models for
  longitudinal and time-to-event data.
\newblock {\em Biometrics}, 67(3):819--829.

\bibitem[Rizopoulos, 2012]{rizopoulos2012joint}
Rizopoulos, D. (2012).
\newblock {\em Joint models for longitudinal and time-to-event data: With
  applications in R}.
\newblock CRC press.

\bibitem[Selvin et~al., 2017]{selvin2017stock}
Selvin, S., Vinayakumar, R., Gopalakrishnan, E., Menon, V.~K., and Soman, K.
  (2017).
\newblock Stock price prediction using lstm, rnn and cnn-sliding window model.
\newblock In {\em 2017 international conference on advances in computing,
  communications and informatics (icacci)}, pages 1643--1647. IEEE.

\bibitem[Spitoni et~al., 2018]{spitoni2018prediction}
Spitoni, C., Lammens, V., and Putter, H. (2018).
\newblock Prediction errors for state occupation and transition probabilities
  in multi-state models.
\newblock {\em Biometrical Journal}, 60(1):34--48.

\bibitem[Staar et~al., 2019]{staar2019anomaly}
Staar, B., L{\"u}tjen, M., and Freitag, M. (2019).
\newblock Anomaly detection with convolutional neural networks for industrial
  surface inspection.
\newblock {\em Procedia CIRP}, 79:484--489.

\bibitem[Topol, 2019]{topol2019high}
Topol, E.~J. (2019).
\newblock High-performance medicine: the convergence of human and artificial
  intelligence.
\newblock {\em Nature medicine}, 25(1):44--56.

\bibitem[Troyanskaya et~al., 2001]{troyanskaya2001missing}
Troyanskaya, O., Cantor, M., Sherlock, G., Brown, P., Hastie, T., Tibshirani,
  R., Botstein, D., and Altman, R.~B. (2001).
\newblock Missing value estimation methods for dna microarrays.
\newblock {\em Bioinformatics}, 17(6):520--525.

\bibitem[van Houwelingen and Putter, 2011]{van2011dynamic}
van Houwelingen, H. and Putter, H. (2011).
\newblock {\em Dynamic prediction in clinical survival analysis}.
\newblock CRC Press.

\bibitem[Van~Houwelingen, 2007]{van2007dynamic}
Van~Houwelingen, H.~C. (2007).
\newblock Dynamic prediction by landmarking in event history analysis.
\newblock {\em Scandinavian Journal of Statistics}, 34(1):70--85.

\bibitem[Vilone and Longo, 2021]{vilone2021notions}
Vilone, G. and Longo, L. (2021).
\newblock Notions of explainability and evaluation approaches for explainable
  artificial intelligence.
\newblock {\em Information Fusion}, 76:89--106.

\bibitem[Vincent et~al., 2009]{mortalityICU}
Vincent, J., Rello, J., and Marshall, J. (2009).
\newblock International study of the prevalence and outcomes of infection in
  intensive care units.
\newblock {\em JAMA}, 302(21):2323--9.

\bibitem[Wilks, 2011]{wilks2011statistical}
Wilks, D.~S. (2011).
\newblock {\em Statistical methods in the atmospheric sciences}, volume 100.
\newblock Academic press.

\bibitem[Zeng et~al., 2022]{zeng2022deep}
Zeng, Z., Hou, Z., Li, T., Deng, L., Hou, J., Huang, X., Li, J., Sun, M., Wang,
  Y., Wu, Q., et~al. (2022).
\newblock A deep learning approach to predicting ventilator parameters for
  mechanically ventilated septic patients.
\newblock {\em arXiv preprint arXiv:2202.10921}.

\bibitem[Zhao and Feng, 2020]{zhao2020deep}
Zhao, L. and Feng, D. (2020).
\newblock Deep neural networks for survival analysis using pseudo values.
\newblock {\em IEEE journal of biomedical and health informatics},
  24(11):3308--3314.

\bibitem[Zheng et~al., 2017]{zheng2017learning}
Zheng, H., Fu, J., Mei, T., and Luo, J. (2017).
\newblock Learning multi-attention convolutional neural network for
  fine-grained image recognition.
\newblock In {\em Proceedings of the IEEE international conference on computer
  vision}, pages 5209--5217.

\end{thebibliography}


\begin{thebibliography}{}

\bibitem[Abramowitz and Stegun, 1964]{abramowitz1964handbook}
Abramowitz, M. and Stegun, I.~A. (1964).
\newblock {\em Handbook of mathematical functions with formulas, graphs, and
  mathematical tables}, volume~55.
\newblock US Government printing office.

\bibitem[Boull{\'e} et~al., 2020]{boulle2020rational}
Boull{\'e}, N., Nakatsukasa, Y., and Townsend, A. (2020).
\newblock Rational neural networks.
\newblock {\em Advances in Neural Information Processing Systems},
  33:14243--14253.

\bibitem[Chen and Qi, 2019]{chen2019representation}
Chen, Y. and Qi, B. (2019).
\newblock Representation learning in intraoperative vital signs for heart
  failure risk prediction.
\newblock {\em BMC medical informatics and decision making}, 19(1):1--15.

\bibitem[Keogh and Pazzani, 2000]{keogh2000scaling}
Keogh, E.~J. and Pazzani, M.~J. (2000).
\newblock Scaling up dynamic time warping for datamining applications.
\newblock In {\em Proceedings of the sixth ACM SIGKDD international conference
  on Knowledge discovery and data mining}, pages 285--289.

\bibitem[Mundhenk et~al., 2019]{mundhenk2019efficient}
Mundhenk, T.~N., Chen, B.~Y., and Friedland, G. (2019).
\newblock Efficient saliency maps for explainable ai.
\newblock {\em arXiv preprint arXiv:1911.11293}.

\bibitem[Platt et~al., 1999]{platt1999probabilistic}
Platt, J. et~al. (1999).
\newblock Probabilistic outputs for support vector machines and comparisons to
  regularized likelihood methods.
\newblock {\em Advances in large margin classifiers}, 10(3):61--74.

\bibitem[Silverman et~al., 1972]{silverman1972special}
Silverman, R.~A. et~al. (1972).
\newblock {\em Special functions and their applications}.
\newblock Courier Corporation.

\bibitem[Simonyan et~al., 2013]{simonyan2013deep}
Simonyan, K., Vedaldi, A., and Zisserman, A. (2013).
\newblock Deep inside convolutional networks: Visualising image classification
  models and saliency maps.
\newblock {\em arXiv preprint arXiv:1312.6034}.

\bibitem[Troyanskaya et~al., 2001]{troyanskaya2001missing}
Troyanskaya, O., Cantor, M., Sherlock, G., Brown, P., Hastie, T., Tibshirani,
  R., Botstein, D., and Altman, R.~B. (2001).
\newblock Missing value estimation methods for dna microarrays.
\newblock {\em Bioinformatics}, 17(6):520--525.

\bibitem[Ye et~al., 2019]{ye2019similarity}
Ye, Y., Jiang, J., Ge, B., Dou, Y., and Yang, K. (2019).
\newblock Similarity measures for time series data classification using grid
  representation and matrix distance.
\newblock {\em Knowledge and Information Systems}, 60(2):1105--1134.

\bibitem[Ypma, 1995]{ypma1995historical}
Ypma, T.~J. (1995).
\newblock Historical development of the newton--raphson method.
\newblock {\em SIAM review}, 37(4):531--551.

\end{thebibliography}

\end{document}


{
\begin{center}
  {\Huge \textbf{Supplementary Material}}
\end{center}
}

\section{Data, covariates and hazards}
\label{icuai:study design} 
This study was conducted within the framework of the Molecular Diagnosis and Risk Stratification of Sepsis (MARS) study (ClinicalTrials.gov identifier NCT01905033), a prospective ICU cohort, for which the institutional review board approved an opt-out method of informed consent (protocol number 10-056C). 
Time-fixed variables included in the model are reported in Table~\ref{tab:baseline_predictors}, while time-dependent covariates are listed in Table~\ref{tab:timedep_predictors}.
{\tiny
\begin{table}[h]
    \centering
    \begin{tabular}{|p{3cm}|p{12cm}|}
    \hline
         \textbf{Variable name} & \textbf{Variable description} \\
    \hline
         Sex & Sex (male/female) \\
    \hline
         Age& Age at ICU admission \\
    \hline
         Immunodeficiency & Immunocompromised status; defined as having acquired immune deficiency syndrome, the use of corticosteroids in high doses (equivalent to prednisolone of $>$75 mg/day for at least 1 week), current use of immunosuppressive drugs, current use of antineoplastic drugs, recent hematologic malignancy, or documented humoral or cellular deficiency \\
    \hline
         Readmission& Previous ICU admission during current hospitalization period \\
    \hline
         Primary specialty& Diagnostic category of ICU admission (cardiovascular, gastrointestinal, neurological, respiratory, post-transplantation, trauma, other)\\
    \hline
         Diabetes Mellitus& Medical history of diabetes mellitus \\
    \hline
    Chronic corticosteroid use & Chronic medication use: systemic corticosteroids\\
    \hline
    Chronic organ failure & Presence of chronic organ insufficiency with one of the following conditions documented in medical history:
    \begin{itemize}
        \item Chronic heart failure defined as the medical history of chronic NYHA class 2-4  or documented ejection fraction $<$45\% (on echography in 2 years prior to ICU admission) or orthopnea with chronic diuretic use
        \item Severe cardiovascular insufficiency defined as angina or dyspnea in rest or during minimal exercise (NYHA IV)
        \item Chronic renal insufficiency defined as chronically elevated serum creatinine $>$177 μmol/L or chronic dialysis
        \item Chronic restrictive, obstructive or vascular pulmonary disease leading to severe functional impairment
        \item Chronic liver failure with portal hypertension (with positive liver biopsy) and/or upper gastrointestinal bleeding due to portal hypertension and/or episode of hepatic encephalopathy/coma due to medical history of liver failure 
    \end{itemize}
     \\
    \hline
    Admission type & Admission to a medical/surgical tertiary ICU \\
    \hline
    \end{tabular}
    \caption{Table with all the baseline predictors.}
    \label{tab:baseline_predictors}
\end{table}
\begin{table}
    \centering
    \begin{tabular}{|p{3cm}|p{12cm}|}
    \hline
         \textbf{Variable name} & \textbf{Variable description} \\
    \hline
         Heart rate & Median of 1-hour mean heart rate (bpm) \\
    \hline
         Blood pressure& Median of 1-hour mean blood pressure, either invasive mean arterial blood pressure measurement or non-invasive cuff (mmHg)\\
    \hline
         Oxygen saturation & Median of 1-hour mean oxygen saturation (\%)\\
    \hline
         Respiratory rate& Median of 1-hour mean respiratory rate (rpm) \\
    \hline
         Pulse& Median of 1-hour mean pulse pressure (difference between systolic and diastolic blood pressure, mmHg)\\
    \hline
         Invasive mechanical ventilation& Last observed mechanical ventilation status \\
    \hline
        FiO2 & Last observed FiO2 (inspired oxygen concentration) value in 8 hours \\
    \hline
    Chronic corticosteroid use & Chronic medication use: systemic corticosteroids\\
    \hline
    Fever & Presence of fever in last 8 hours ($>$38 degrees Celsius)
     \\
    \hline
    Fluid balance & Fluid balance (mL) over the past 8 hours \\
    \hline
    Urine output & Total urine output (mL) in 8-hour window\\
    \hline
    Suctioned sputum & Total number of times sputum was suctioned and observed within an 8-hour time window \\
    \hline
    Worsening CNS status & Either decrease in consciousness (either a decrease in GSC M-score or worsening RASS score) or onset of new delirium episode in the past 8 hours\\
    \hline
    CRP (last value) & Last observed CRP (mg/L)\\
    \hline
    CRP (change) & Unit change in CRP relative to CRP 24 hours earlier (mg/L)\\
    \hline
    White blood cell count (last value) & Last observed white blood cell count (x$10^9$/L)\\
    \hline
    White blood cell count (change) & Unit change in white blood cell (WBC) count relative to WBC hours earlier (x$10^9$/L)\\
    \hline
    Platelet count (last value) & Last observed platelet count (x$10^9$/L)\\
    \hline
    Platelet count (change) & Unit change in platelet count relative to platelet count 24 hours earlier (x$10^9$/L)\\
    \hline
    Prothrombin time (last value) & Last observed prothrombin time (seconds)\\
    \hline
    Creatinine (last value) & Last observed creatinine (\mu mol/L)\\
    \hline
    Creatinine (change) & Unit change in creatinine relative to creatinine 24 hours earlier  (\mu mol/L)\\
    \hline
    Total bilirubin (last value) & Last observed total bilirubin  (\mu mol/L)\\
    \hline
    Total bilirubin (change) & Unit change in total bilirubin relative to bilirubin 24 hours earlier (\mu mol/L)\\
    \hline
    Bicarbonate (change) & Unit change of bicarbonate relative to bicarbonate 24 hours earlier (mmol/L)\\
    \hline
    Lactate (last value) & Last observed lactate (mmol/L)\\
    \hline
    Increase in vasopressor rate & Increase in mean norepinephrine dose relative to previous 8-h window\\
    \hline
    Increase in insulin dose & Increase in mean insulin dose relative to previous 8-h window\\
    \hline
    Gram+ in respiratory culture & Gram-positive bacteria cultured in the airway (the result of the most recent culture)\\
    \hline
    Candida in respiratory culture & Candida species cultured in the airway (the result of the most recent culture)\\
    \hline
    \end{tabular}
    \caption{Table with all the time-dependent predictors.}
    \label{tab:timedep_predictors}
\end{table}
}
\\
The cause-specific hazards for infection $\beta_1^{(0)}$ of the fitted Deep LM-CR model are shown in Table~\ref{tab:case_specific_24_hour}.

\begin{table}
    \centering
    \begin{tabular}{|c|c|c|}
        \hline
         \textbf{Covariate} &  $\boldsymbol{\beta}^{(0)}_1$ & $\boldsymbol{\beta}$-CI\\
         \hline
         Urine output & 1 & 0.99-1.01\\
         \hline
         Suctioned sputum & 1 & 0.99-1.02\\
         \hline
         SpO2 & 0.97 & 0.95-0.98\\
         \hline
         Respiratory Rate & 1 & 0.99-1.01\\
         \hline
         Readmission & 0.92 & 0.85-1.11\\
         \hline
         Pulse pressure & 1 & 0.99-1.01\\
         \hline
         \rowcolor{lightgray}
         Gender (male) & 1.4 & 1.20-1.53\\
         \hline
         \rowcolor{lightgray}
         Increase in vasopressor rate & 1.4& 1.25-1.49\\
         \hline
         Mechanical ventilation & 1 & 0.88-1.21\\
         \hline
         mean Blood Pressure & 1 & 0.99-1.01\\
         \hline
         White Blood Cells (last value)& 1 & 0.99-1.01\\
         \hline
         Prothrombin time (last value) & 1 & 0.99-1.01\\
         \hline
         Platelets (last value) & 1 & 0.99-1.01\\
         \hline
         Lactose (last value) & 0.98 & 0.95-1.04\\
         \hline
         CRP (last value) & 1 & 0.99-1.01\\
         \hline
         Bilirubin (total) & 1 & 0.99-1.01\\
         \hline
         Increase in insulin dose (total) & 1 & 0.99-1.01\\
         \hline
         \rowcolor{lightgray}
         Immune compromise (total) & 1.2 & 1.01-1.54\\
         \hline
         Heart Rate (total) & 1 & 0.95-1.05\\
         \hline
         \rowcolor{lightgray}
         Gram+ respiratory culture (total) & 1.1  & 1.01-1.54\\
         \hline
         Fluid balance & 1 & 0.99-1.01\\
         \hline
         FiO2 (total) & 1 & 0.99-1.01\\
         \hline
         Worsening CNS status (total) & 1.1 & 1.02-1.18\\
         \hline
         \rowcolor{lightgray}
         Fever  & 2 & 1.86-2.75\\
         \hline
         APACHE-Trauma & 1 & 0.92-1.34\\
         \hline
         APACHE-Transp & 0.97 & 0.86-1.26\\
         \hline
         APACHE-Respir & 0.67 & 0.54-0.77\\
         \hline
         APACHE-Other & 0.58 & 0.33-0.96\\
         \hline
         \rowcolor{lightgray}
         APACHE-Neuro & 1.2 & 1.01-1.48\\
         \hline
         APACHE-Gastro & 0.71 & 0.54-0.96\\
         \hline
         Admission Type (surgical) & 1.3 & 1.16-1.44\\
         \hline
         Diabetes mellitus & 0.82 & 0.73-0.99\\
         \hline
         White Blood cells (change) & 1 & 0.99-1.01\\
         \hline
         Platelets (change) & 1 & 0.99-1.01\\
         \hline
         pCO2 (change) & 1 & 0.99-1.01\\
         \hline
         CRP (change) & 1 & 0.99-1.01\\
         \hline
         Bilirubine total (change) & 1 & 0.99-1.01\\
         \hline
         Bicarbonate (change) & 1 & 0.99-1.01\\
         \hline
         Corticosteroids & 1.2 & 0.99-1.48\\
         \hline
         \rowcolor{lightgray}
         CNN Risk Score & 4.8 & 3.05-6.72\\
         \hline
          Chronic organ failure & 1.1 & 0.95-1.28\\
         \hline
         Candida respiratory culture & 0.82 & 0.72-0.91\\
         \hline
         age & 1 & 0.99-1.01\\
         \hline
    \end{tabular}
    \caption{Cause-specific hazards  of ICU-AI}
    \label{tab:case_specific_24_hour}
\end{table}

\section{ANN for high-frequency vital signs}\label{icuai:model_selection}
In this section, we present the model selection of the best ANN design for high-frequency vital signs, i.e., the five EHRs used to compound the risk score of infection.
In addition to this, we also considered two classical \ac{ML} models such as the \ac{LR} and the \ac{SVM}.
The model selection was accomplished considering two different prediction windows: 24 hours and 48 hours.
To determine the best model, we made a search over a fine grid of hyper-parameters; according to the set of hyper-parameters selected, we proposed the best model that had the highest accuracy.
In this phase, accuracy is evaluated as the averaged AUROC of a  5-fold cross-validated model.

\subsection{Model selection}
We first tested the level of accuracy of \ac{LR}, \ac{SVM},  and \ac{MLP}, after aggregating the time series along the time domain.
Thus, simple statistics of the time-series instances were extracted along the whole 24-hour (or 48-hour) domain of the time-series instances (e.g., mean value, standard deviation, skewness, kurtosis, minimum, and maximum value); this operation was accomplished for each one of the five vital signs.
The total of input features we obtained was 31.
It is important to mention that these extracted input features were standardized (mean value equal to zero and standard deviation equal to the unit).

\begin{figure}
    \begin{subfigure}{.5\textwidth}
    \centering
        \includegraphics[width= 1\linewidth]{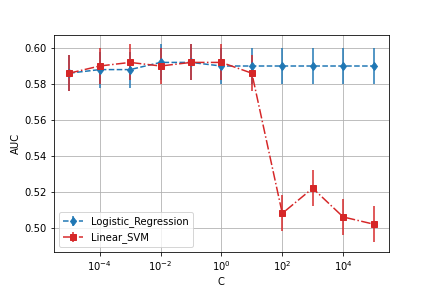}
        \caption{}
    \end{subfigure}
    \begin{subfigure}{.5\textwidth}
    \centering
         \includegraphics[width= 1\linewidth]{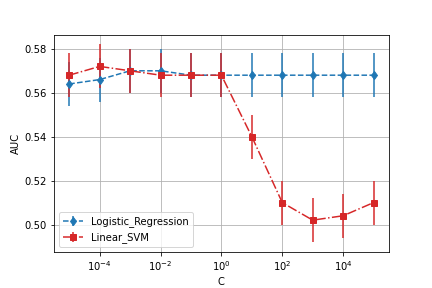}
         \caption{}
    \end{subfigure}
    \caption{\ac{AUROC} as a function of the inverse shrinkage parameter ($C$) for (a) the 24-hour and (b) the 48-hour prediction model.
    The red curve concerns the \ac{SVM} model, while the blue line the \ac{LR} model.}
    \label{fig:model_sel_LR_SVM}
\end{figure}

Regularization was included in both \ac{LR} and the \ac{SVM}; these models were penalized with the $L^{2}$ norm of the weights. 
We used the inverse of the shrinkage parameter (here denoted as $C$) which represented the unique hyper-parameter of these two models. 
Accordingly, we searched for the best $C$ optimizing the \ac{AUROC}; results in Figure~\ref{fig:model_sel_LR_SVM}.
We observed, however, that both models cannot achieve an \ac{AUROC} score larger than 0.59.  
Similar low results were found for 24-hour and 48-hour models.

\begin{figure}
    \begin{subfigure}{.5\textwidth}
         \centering
        \includegraphics[width= 1\linewidth]{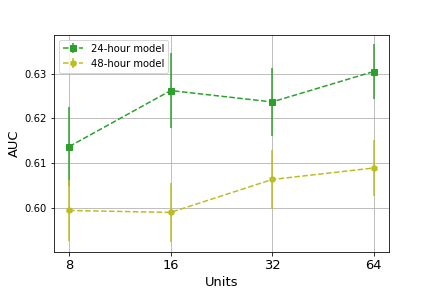}
        \caption{}
    \end{subfigure}
    \begin{subfigure}{.5\textwidth}
         \centering
         \includegraphics[width= 1\linewidth]{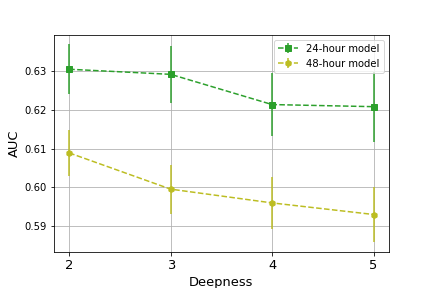}
         \caption{}
    \end{subfigure}
    \begin{subfigure}{.5\textwidth}
         \centering
         \includegraphics[width= 1\linewidth]{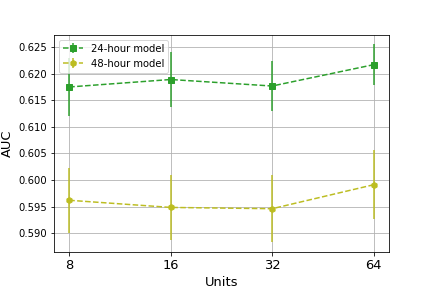}
         \caption{}
    \end{subfigure}
    \begin{subfigure}{.5\textwidth}
         \centering
         \includegraphics[width= 1\linewidth]{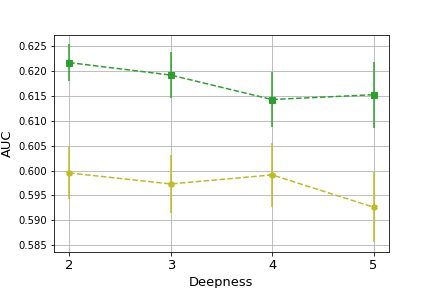}
         \caption{}
    \end{subfigure}
    \caption{\ac{MLP} model. Maximal \ac{AUROC} as a function of the hyper-parameters \emph{Units} (a.k.a, number of units) and \emph{Deepness}. 
    Each plot presents the behavior of both the 24-hour (green line) and 48-hour models (yellow line). The following cases are considered:
    (a) Number of units and ReLU activation function,
    (b) Deepness and ReLU activation function,
    (c) Number of units and tanh activation function,
    (d) Deepness and tanh activation function.}
    \label{fig:model_sel_MLP}
\end{figure}

Unlike the previous models, when considering the \ac{MLP}, the best accuracy of the model is intimately connected with the search for a wider set of optimal hyper-parameters.
One has indeed more parameters to tune:  \emph{number of units, deepness, dropout rate, learning rate, activation function}, and the \emph{batch size}.
As mentioned, the tuning of these hyper-parameters was accomplished after inquiring about the set of parameters returning the maximal averaged \ac{AUROC} in a 5-fold cross-validation.
Note that the visualization of the curve of the \ac{AUROC} score as a function of the hyper-parameters is quite impractical, because of the large number of parameters we had to tune.
Instead, we regarded all the configurations after constraining one hyper-parameter (e.g., all configurations with deepness equal to 3 or a number of units equal to 16), and then we took that one with the highest \ac{AUROC} score (we shall refer to this as the \emph{maximal \ac{AUROC} of a hyper-parameter}). 
Then, we represented the maximal \ac{AUROC} to get an idea about the variation of \ac{AUROC} with respect to one single hyper-parameter.
We utilized an \ac{MLP} composed of a sequence of dense layers only (we refer to the number of dense layers as \emph{deepness}) equipped with a non-linear activation function; each one of these layers is regularized with a dropout layer placed just underneath it.
The \ac{MLP} model was designed to optimize the binary cross-entropy by means of the ADAM optimizer (hyper-parameters such as \emph{learning rate} and \emph{batch size} are functionalities of the optimizer).
In Figure~\ref{fig:model_sel_MLP}a and \ref{fig:model_sel_MLP}b are shown the \ac{MLP} models with a ReLU activation function (i.e., $\operatorname{ReLU}(x) = \max{(0, x)}$): the model did not achieve an \ac{AUROC} score higher than 0.63.
Similarly, the choice of a hyperbolic tangent (tanh) activation function returned a similar result; see Figure~\ref{fig:model_sel_MLP}c and \ref{fig:model_sel_MLP}d.

The next class of models we analyzed are the \ac{CNN}s.
Similarly to the \ac{MLP}, we adopted the same strategy of optimizing the \ac{AUROC} over a set of hyper-parameters: the \emph{number of convolutional filters, kernel size, deepness, learning rate, dropout rate}, and \emph{the batch size}.
In this case, the activation function has not been included in the hyper-parameters to tune; unlike the \ac{MLP} model, we only considered the ReLU activation function.
The reason for such a decision comes after noting that several tests with a \emph{one held-out} approach revealed that sigmoidal activation functions (e.g., sigmoid or hyperbolic tangent) affected the predictive power of the model; we always obtained an AUROC lower than 0.60 for different combinations of \emph{power} (i.e., the number of filters $times$ dropout rate), \emph{deepness} (i.e., number of hidden layers), and \emph{receptive field} (i.e., the combination of kernel and max-pooling layers of different size).

Before propagating the vital signs through the CNN model, a few pre-processing steps had to be performed.
Firstly, one linear transformation was applied to all the time-series instances to give a more compact representation in the range [-1, 1].
Specifically, we applied the same-type transformation to all instances; according to the type of vital sign to rescale, a precise linear transformation was applied.
For example, we used the same linear transformation to rescale all heart rate time series contained in all the time series instances; while for all other vital signs (e.g., breath rate or others), we developed and used a different one.
Hence, according to the overall statistics of all the vital signs available to us, for each time-series feature, we constructed a linear map that make a rescaling in the domain [-1, 1]. 
For example, we can consider the heart rate predictor from the time-series instance; from the overall statistics of all heart rate signs of all patients, we recorded a global minimum and maximum equal to 41 and 239 beats per minute, respectively.
Accordingly, denoting with $X^{HR}_i(t)$ the heart rate feature of the $i$-th time series instance, we applied to $X^{HR}_i(t)$ the following transformation
\begin{equation*}
    X^{HR}_i(t) \rightarrow \frac{2X^{HR}_i(t) -41 \text{bpm} -239 \text{bpm}}{239\text{bpm} - 41\text{bpm}}.
\end{equation*}

Unlike standardization (i.e., one imposes that all time-series features have unitary variance and zero mean value), the application of these data-based linear transformations does not drastically distort proper characteristics of the vital signals such as scale (i.e., the mean value) and energy (i.e., the empirical second moment) values.
In addition, data were processed using the Piecewise Approximate Aggregation (\ac{PPA}) method \cite[]{keogh2000scaling}. 
Instead of representing all high-scale details, we obtained a reduced but informative representation of vital signals while maintaining the lower bound of distance measurements in Euclidean space\cite[]{chen2019representation}. Therefore, we used \ac{PPA} to aggregate time intervals of length 9 minutes.

\begin{figure}
    \centering
    \begin{subfigure}{.49\textwidth}
        \includegraphics[width= 1\linewidth]{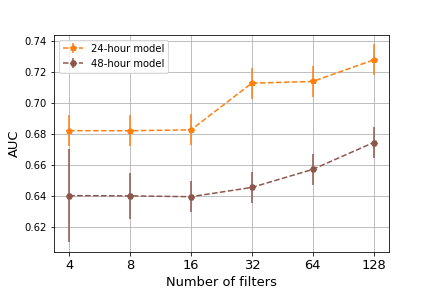}
        \caption{}
    \end{subfigure}
    \begin{subfigure}{.49\textwidth}
         \includegraphics[width= 1\linewidth]{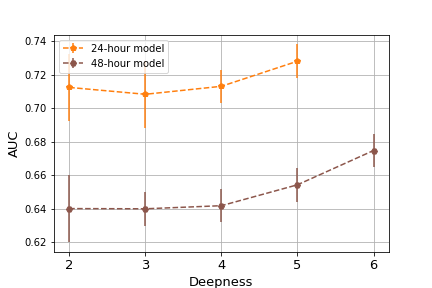}
         \caption{}
    \end{subfigure}
    \begin{subfigure}{.49\textwidth}
         \includegraphics[width= 1\linewidth]{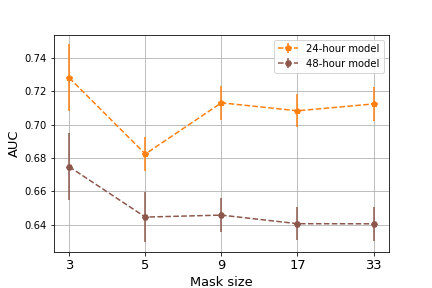}
         \caption{}
    \end{subfigure}
    \caption{\ac{CNN} model. Maximal AUROC as a function of the hyper-parameters \emph{number of filters} (a), \emph{deepness} (b), and \emph{kernel size}. 
    Each plot presents the behavior of both the 24-hour and 48-hour models. }
    \label{fig:model_sel_1DCNN}
\end{figure}

In Figure~\ref{fig:model_sel_1DCNN}a, we see that the accuracy of the CNN increases with the number of filters:
a high number of filters, such as 128, makes both 24-hour and 48-hour models accurate with AUROC 0.72 and 0.67, respectively.
We first investigated the \ac{1-D CNN}.
The composition of many hidden layers is another key feature of enabling the model to be performative.
In fact, in Figure~\ref{fig:model_sel_1DCNN}b we can observe how 5 layers are enough for the 24-hour model (AUROC in the range [0.70, 0.72]), whereas 6 convolutional layers were necessary to enable the 48-hour model to achieve the highest AUROC (0.67). 
Conversely, the amplitude of the convolutional masks reduces the AUROC values, especially when masks of size 17 or 33 were considered; see  Figure~\ref{fig:model_sel_1DCNN}c; convolutional masks of size 3 enable keeping AUROC values 0.72 and 0.67 for both the 24-hour and 48-hour models.
Thus, our investigation for the best configuration, revealed that powerful (i.e., with many filters) and deep networks with small-sized kernels are the types of CNN models to use.

\begin{figure}
    \centering
    \includegraphics[width= .66\textwidth]{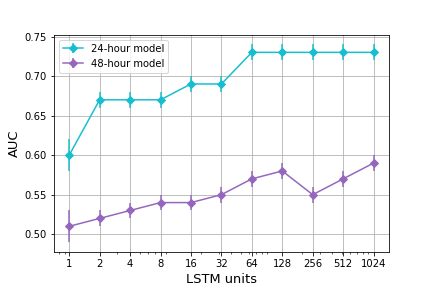}
    \caption{CNN LSTM model. \ac{AUROC} as a function of the number of units of the \ac{LSTM} layer. 
    The choice of all other hyperparameters of the CNN-LSTM model is identical to the ones of the optimal \ac{CNN} model.}
    \label{fig:model_sel_CNN_LSTM}
\end{figure}

We compared a pure convolutional approach (i.e., the 1-D CNN model) with a CNN-LSTM model.
That is a CNN model with the exact architecture of the best 1-D CNN we found, except for the fact that an LSTM layer replaces the \emph{flatten layer} of the \ac{CNN} model.
The presence of an LSTM layer has the scope of giving an array-structured feature map while analyzing the connections of input feature maps along the time domain. 
Also, the LSTM layer possesses only one relevant hyper-parameter, i.e., the \emph{number of units} denoting the dimensionality of the array-structured encoding.
Thus, without starting over with a new fine grid-based search, we only focused on the possibility of 
ameliorating the best 1-D CNN we found with the replacement of the \emph{flatten layer} with the LSTM layer.
Still, the parameter \emph{number of units} is the unique variable we studied. 
In Figure~\ref{fig:model_sel_CNN_LSTM}, we can see that an increase in the number of units of the LSTM layer did not return a net increase in the \ac{AUROC} score.
For the 24-hour model, the plateau region starting at unit 64 reveals that the CNN-LSTM model is as accurate as the CNN model, i.e., AUROC score equal to $0.72\pm 0.01$.
The 48-hour model cannot achieve \ac{AUROC} values larger than 0.6, given any configuration of the LSTM units. 
{}

\begin{figure}
    \centering
    \includegraphics[width= .99\textwidth]{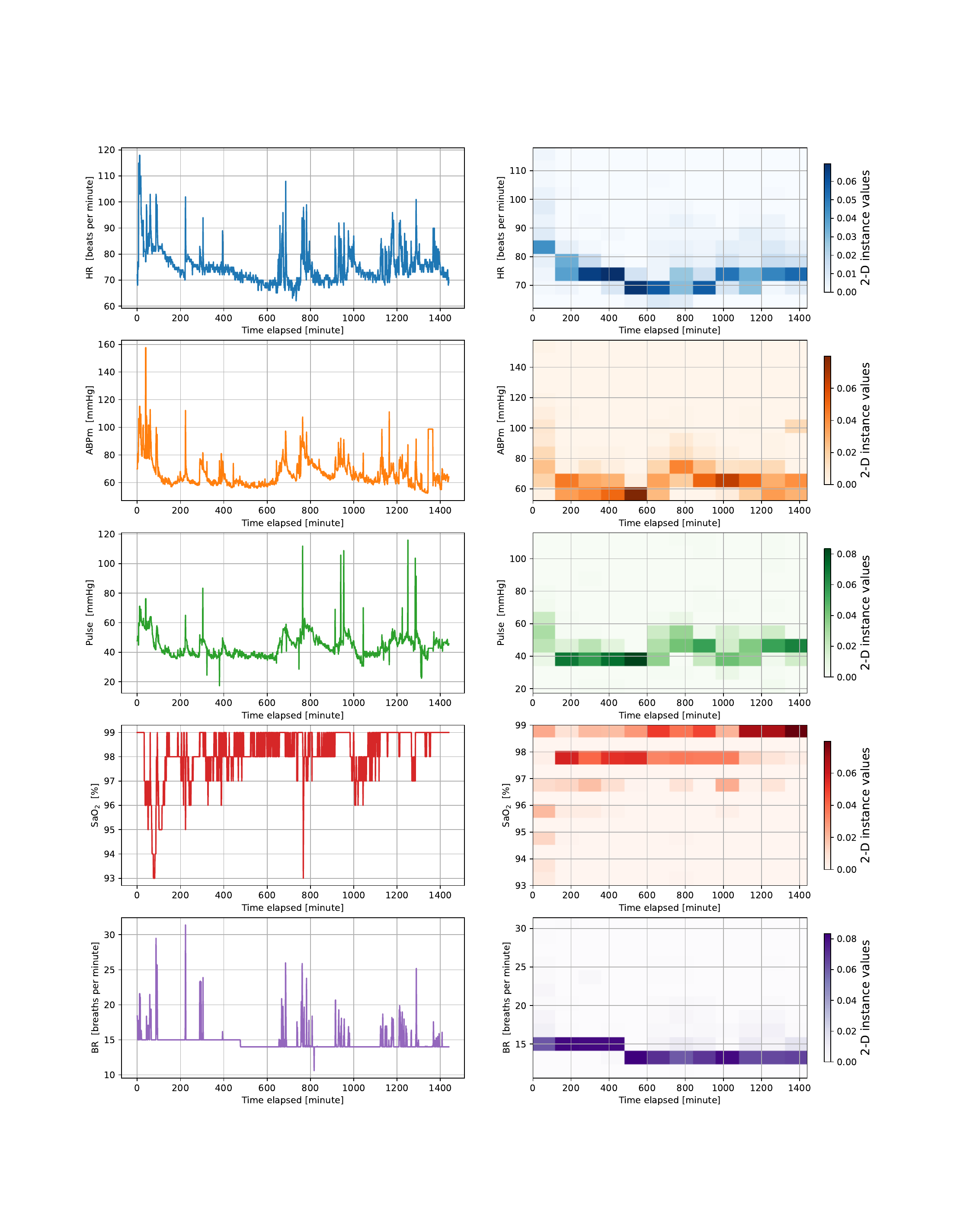}
    \caption{Example of 2-D representation of a time-series instance. On the left column the time-series features (\ac{EHR}), while on the right columns their 2-D representation}
    \label{fig:Example_2D_instance}
\end{figure}

The last class of CNN models that we tested is the 2-D CNN. 
Although 1-D convolutional layers represent the most natural choice for the analysis of Time-Series structured data, a two-dimensional convolutional-based approach is always possible if one provides a 2-D representation of the sequential data.
For example, the method developed by \cite{ye2019similarity} offers the possibility of giving a 2-D representation of time series data, i.e. a 2-D binning is performed, where each 2-D bin counts the number of records falling in a specific range of values and at some precise moments along the time domain; see the example in Figure~\ref{fig:Example_2D_instance}.
A 2-D grid-structured representation of the \ac{EHR} enabled us to investigate the possibility of using a \ac{2-D CNN} model to identify the onset of \ac{ICU-AI}.

\begin{figure}
    \centering
    \begin{subfigure}{.49\textwidth}
        \includegraphics[width= 1\linewidth]{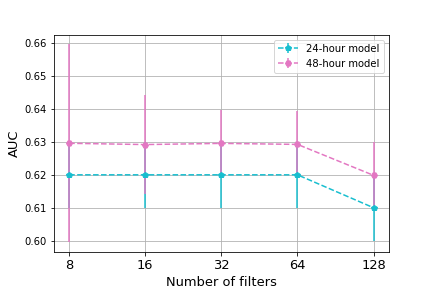}
        \caption{}
    \end{subfigure}
    \begin{subfigure}{.49\textwidth}
         \includegraphics[width= 1\linewidth]{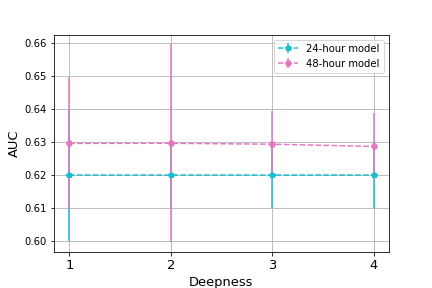}
         \caption{}
    \end{subfigure}
    \begin{subfigure}{.49\textwidth}
         \includegraphics[width= 1\linewidth]{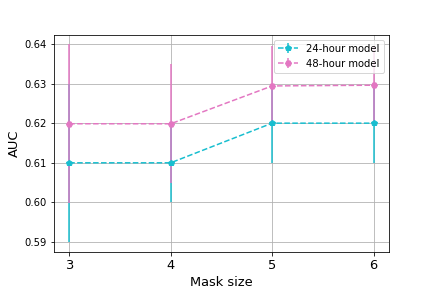}
         \caption{}
    \end{subfigure}
    \begin{subfigure}{.49\textwidth}
         \includegraphics[width= 1\linewidth]{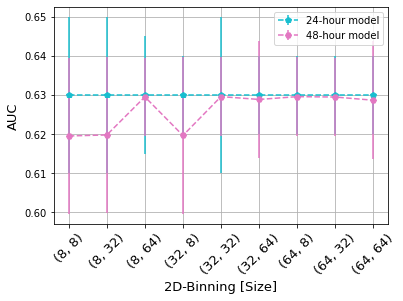}
         \caption{}
    \end{subfigure}
    \caption{\ac{2-D CNN} model. Maximal AUROC as a function of the hyper-parameters \emph{number of filters} (a), \emph{deepness} (b), \emph{kernel size} (c), and the \emph{dimensionality of the 2-D bins} (d). 
    Each plot presents the behaviour of both the 24-hour (cyan) and 48-hour (pink) models. }
    \label{fig:model_sel_2DCNN}
\end{figure}

Despite sharing a similar structure with the 1-D \ac{CNN}, the implementation of the \ac{2-D CNN} required tuning a larger number of hyper-parameters.
Such an increase in hyper-parameters is mainly due to the 2-D structure of data; unlike the 1-D case, in the 2-D case, both the width and the height of the convolutional masks need to be optimized as well as the height and the width of the 2-D bins representing each time-series feature. 
As with the \ac{1-D CNN} model, we optimally tuned the model on a fine grid of parameters: \emph{number of filters, kernel size (on both the 2 dimensions), deepness, dropout, learning rate, batch size}, and both \emph{width and height of the 2-D bins}.
In Figure~\ref{fig:model_sel_2DCNN}, one can see that drastic changes in the architecture of the \ac{2-D CNN} model do not cause relevant changes in the evaluation of the maximal \ac{AUROC} score.
That is, opting for several configurations in the number of filters (see figure \ref{fig:model_sel_2DCNN}a), in the deepness (see Figure~\ref{fig:model_sel_2DCNN}b), in the size of the convolutional masks (Figure~\ref{fig:model_sel_2DCNN}c),
 and in the height and width of the 2-D bins (Figure~\ref{fig:model_sel_2DCNN}d) do not lead both the 24-hour and the 48-hour models to achieve \ac{AUROC} scores larger than 0.63.

\subsection{Choosing the method for imputation}
\begin{figure}[h]
    \centering
    \begin{subfigure}{.49\textwidth}
        \includegraphics[width= 1\linewidth]{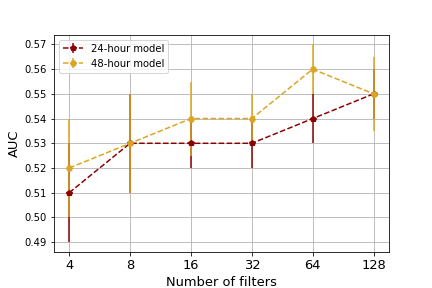}
        \caption{}
    \end{subfigure}
    \begin{subfigure}{.49\textwidth}
         \includegraphics[width= 1\linewidth]{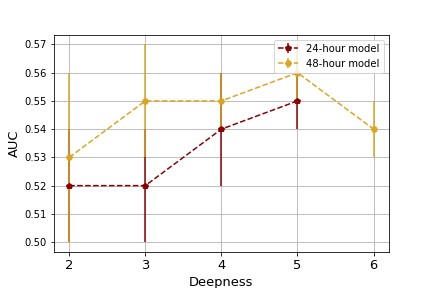}
         \caption{}
    \end{subfigure}
    \begin{subfigure}{.49\textwidth}
         \includegraphics[width= 1\linewidth]{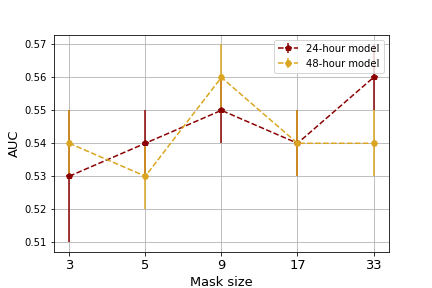}
         \caption{}
    \end{subfigure}
    \caption{\ac{1-D CNN} model fed by \ac{EHR} data imputed via the ICUAI-Imputation method. Maximal AUROC as a function of the hyper-parameters of the \ac{1-D CNN} model.
    (a) \emph{number of filters}, 
    (b)\emph{deepness} (b), 
    (c) \emph{kernel size}.
    Each plot presents the behavior of both the 24-hour and 48-hour models. }
    \label{fig:model_sel_1DCNN_OCM}
\end{figure}

\begin{figure}[h]
    \centering
    \begin{subfigure}{.49\textwidth}
        \includegraphics[width= 1\linewidth]{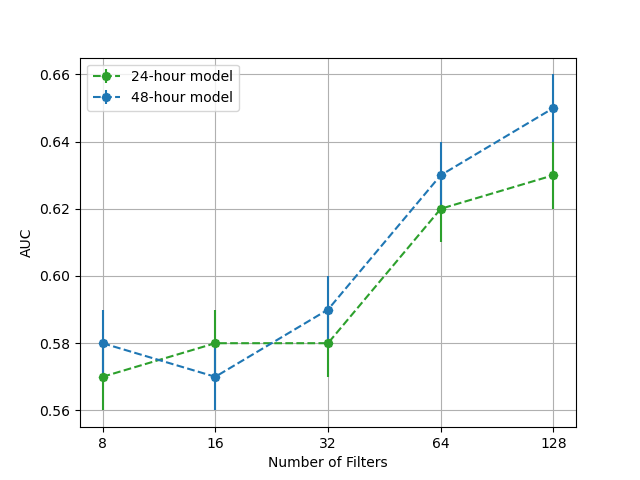}
        \caption{}
    \end{subfigure}
    \begin{subfigure}{.49\textwidth}
         \includegraphics[width= 1\linewidth]{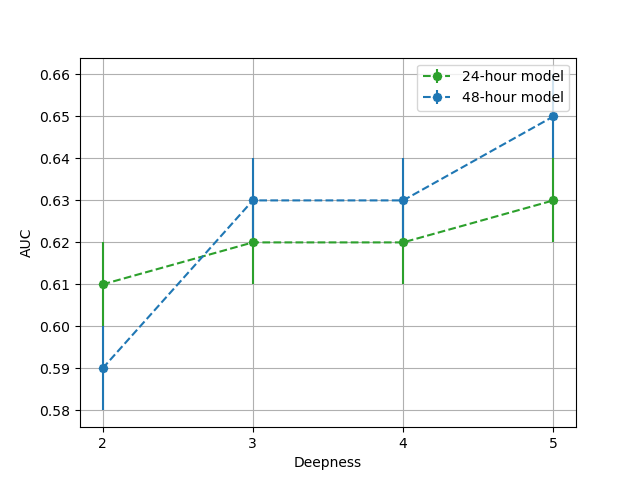}
         \caption{}
    \end{subfigure}
    \begin{subfigure}{.49\textwidth}
         \includegraphics[width= 1\linewidth]{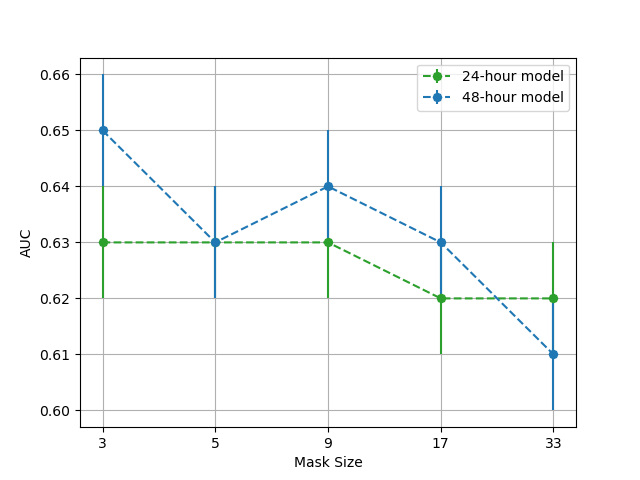}
         \caption{}
    \end{subfigure}
    \caption{\ac{1-D CNN} model fed by \ac{EHR} data imputed via the kNN imputation method. Maximal AUROC as a function of the hyper-parameters of the \ac{1-D CNN} model.
    (a) \emph{number of filters}, 
    (b) \emph{deepness}, 
    (c) \emph{kernel size}.
    Each plot presents the behavior of both the 24-hour and 48-hour models. }
    \label{fig:model_sel_1DCNN_KNN}
\end{figure}

To give quantitative reasoning about the choice of the LOCF imputation method, we present here a comparative analysis involving two alternative methods.
The first method employed was the multivariate kNN (k Nearest Neighbours)\cite[]{troyanskaya2001missing}.
The kNN method, originally designed for matrix data, was here readapted adaptation by flattening the time-series features by organizing the tensor data (number of samples, length of time series, time-series features) into matrix data (number of samples, features). 
Consequently, the features of the latter encapsulated information from each time series at every time point. Subsequently, the kNN method was employed to impute missing data, leveraging the learned connectivity shared by each record with all others within a single instance.
The second method, hereafter referred to as the \emph{ICUAI-Imputation} method, was devised to retrieve information primarily based on nature of missing values under a clinical point of view. 
In essence, missing intervals with an amplitude exceeding 4 hours were substituted with a constant out-of-range value (e.g., 100), while intervals shorter than 4 hours were imputed with a null constant value. 
The rationale behind the ICUAI-imputation method drew inspiration from a practical medical perspective in managing Electronic Health Records (EHR). 
Intervals of approximately 4 hours or longer typically corresponded to the duration of surgical operations, while shorter intervals were often associated with the temporary interruption of ICU monitoring, resulting from the unintentional detachment of devices, whether by a patient or due to device malpositioning.
With this method, we systematically filled specific types of missing intervals with designated placeholder values; this way, the imputed patterns of missing intervals could also include distinct clinical events of interest occurring during the ICU stay. 
As with the model's selection, an investigation over a fine-grid of hyperparameters was conducted. 
In particular, we show results for three of the most determining parameters, i.e., Number of filters, kernel size (or mask size) and deepness. 
In Figures \ref{fig:model_sel_1DCNN_OCM} and \ref{fig:model_sel_1DCNN_KNN} the maximal AUROC is shown. 
We emphasise that the examination of the impact of all the imputation methods on the CNN predictions was conducted through the application of two distinct amplitudes of prediction windows: 24 and 48 hours.
We opted for such a strategy because we aimed to quantify whether the imputation methods considered could actually be beneficial when analysing the longitudinal evolution of data at different time scales.
From Table \ref{tab:imputation_auroc} we observed that the LOCF method emerged as the most promising for both the time scales considered, i.e., 24 and 48 hours (In table \ref{tab:imputation_auroc} we considered models for 24-hour and 48-hour).
The discrepancy with respect to the other methods is net.
For the 24-hour model, the kNN and the ICUAI-Imputation method can't achieve accuracy larger than the ones obtained for the the LOCF imputation method.
This result identifies the LOCF method as the preferred choice for data imputation.
\begin{table}[h]
    \centering
    \begin{tabular}{|l|c|c|}
    \hline
       \textbf{Imputation method}  &  \textbf{AUROC (24-hour model)} &  \textbf{AUROC (48-hour model)}\\
       \hline
       LOCF  & 0.72 ± 0.01 &  0.68 ± 0.01\\
       kNN  & 0.63 ± 0.01 & 0.65 ± 0.01 \\
       ICUAI-Imputation  & 0.56 ± 0.01& 0.56 ± 0.01\\
       \hline
    \end{tabular}
    \caption{Summary of Imputation methods: The highest performance achieved during the validation phase, measured by AUROC, is reported for each imputation method considered. The columns displaying AUROC scores represent either the 24-hour instance model or the 48-hour instance model. AUROC scores have been rounded to the nearest second decimal. Errors were assessed using the Standard Error Mean, and if too short, they were substituted with the minimum error, i.e., 0.01.}
    \label{tab:imputation_auroc}
\end{table}

\subsection{Chosen design} \label{sub:design}
Among all the models considered, the 1-D \ac{CNN} model was not the one with the highest predictive performance. 
Although the 24-hour CNN-LSTM model could be slightly more accurate than the \ac{CNN}, we observed that the latter showed more precise predictive performances even with 48-hour Time-Series instances.
The difference in terms of \ac{AUROC} between both models is marginal for the 24-hour model but instead evident for the 48-hours models.
Moreover, we opted for a 1-D \ac{CNN} model also because our work is aimed by the intent of explaining the activity of pattern recognition via a robust \ac{XAI} method such as the \ac{SMOE} scale (see Section~\ref{sec:smoe}).

For the 24-hour model, we propose the following optimal architecture:
\begin{enumerate}
    \item \emph{Convolutional Layers}: the number of filters on each layer is 128, and each filter has a size of 3 (pixels).
    The result of these convolutions is referred to as \emph{feature maps}.

    \item \emph{Activation Layer}:
    the ReLU function is applied after each convolution operator. 
    This application of a non-linear activation function on the feature maps gives birth to the \emph{activated feature maps}.

    \item \emph{Max-pooling layer}: 
    the activated feature maps are resampled via a Max-pooling operator with a pooling size of 2.
\end{enumerate}
This sequence of hidden layers is repeated five times.
The architecture also encloses a \emph{Dropout layer} after each Max-Pooling layer.
The Dropout layer has a dropout rate of 0.25.
The last feature map is flattened into an array and then propagated in a \emph{fully-connected} layer (\emph{dense layer}) with a sigmoid activation function. 
The activation function returns a positive output between 0 and 1, that is, the risk score denoting the chance of a patient developing an \ac{ICU-AI} episode.
As usual, the loss function is the binary cross-entropy, and the optimizer is the ADAM algorithm.
For the 48-hour model, the architecture is identical to the 24-hour one, except for the fact that the sequence of convolutional and max-poling layers is repeated 6 times.


\section{ANN-based predictions for ICU-AI}
In addition to the two-step strategy presented in our work, we consider a full ANN-based approach.
We would like to understand how the predictive power of a full Deep-learning model can help to forecast the occurrence of acquired infections and then compare these two approaches. 

Thus, in order to make the ANN-based predictions comparable with those obtained by the (Deep) LM-CR, we opted for fitting, at each landmark time, an ANN model fed simultaneously with all kinds of data (i.e., high-frequency, low-frequency, and baseline data) of all patients who are still at risk of experiencing an infectious episode.
As for the (Deep) LM-CR, we focused on the first infectious episode only.
Also, we trained the ANN model with a lead time of 24 hours; namely, if we suppose to train this ANN with the data available at landmark time $\tau$, the response variable will be a dichotomous variable flagging the occurrence of acquired infection at time $\tau + 24\text{ hours}.$
As mentioned, this ANN model was devised to be fed simultaneously with different data types such as time series (high-frequency data) and data matrices (low-frequency and baseline data).
To accommodate this peculiarity of the dataset, we chose to develop a \emph{two-branch artificial neural network} (TB-ANN).

The architecture of this model is therefore composed of two distinct branches each one of them analyzing either the time series or the data matrices; their latent representations are finally conveyed into one prediction layer.
The branch for the analysis of time series is pretty similar to the 1-D CNN developed for the CNN risk score:
\begin{enumerate}
    \item The input time series are processed through a sequence of hidden layers, i.e., Convolutional layer, activation function, max-pooling, and dropout layer. 
    The convolution layer has 128 filters with a kernel size of 3, the max-pooling size is equal to 2, and the dropout rate is equal to 0.2.
    The activation function is a rational activation function \cite[]{boulle2020rational}.
    \item The sequence of hidden layers is then repeated 3 times.
    \item The feature maps are processed through an LSTM layer with 32 output units which represent the \emph{latent representation of the time series}.
\end{enumerate}
Similarly, the branch operating with the data matrices is an MLP with 3 dense layers with 32 output units; the activation function is still a Rational Activation Function. 
Thus, after propagating the data matrices through this sequence of dense layers, \emph{the latent representation of the data matrices} is obtained as an array of dimension 32. 
\begin{figure}[h]
    \centering
    \includegraphics[width= \textwidth]{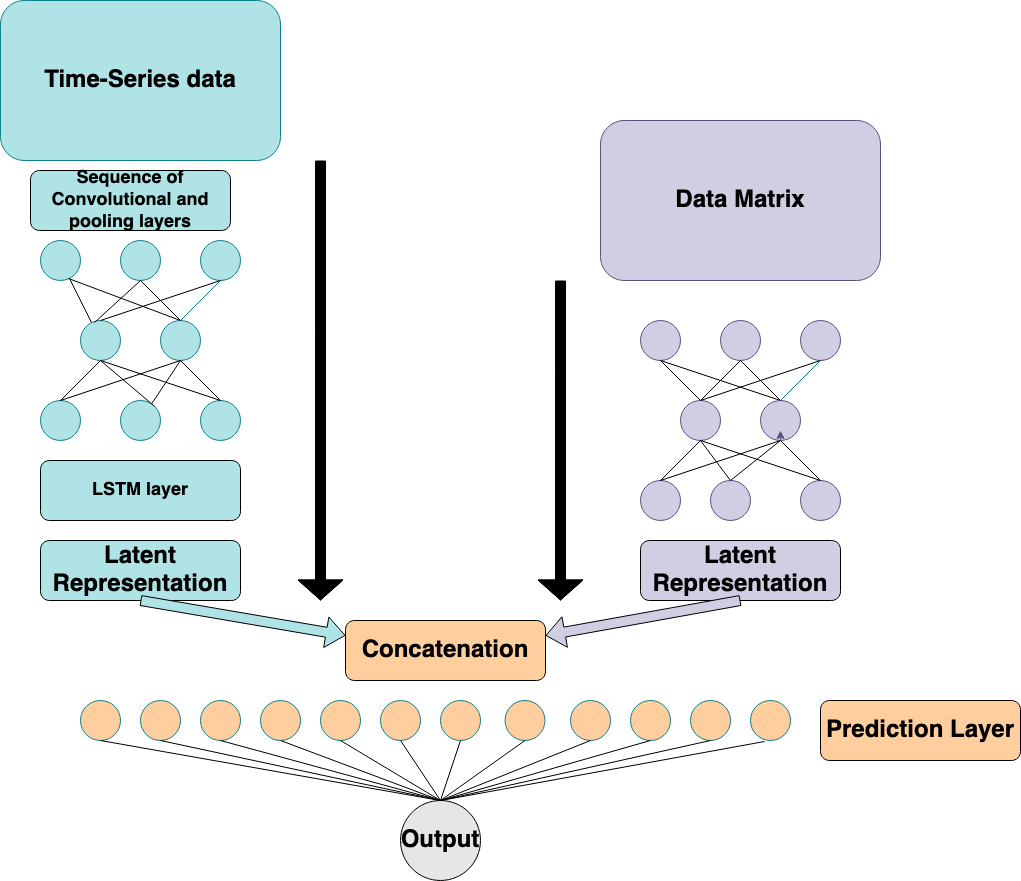}
    \caption{Schematic illustration of the TB-ANN. 
    The right branch (light-blue) is dedicated to encoding the Time-Sereis features; similar to a 1-D CNN, it is made of an ordered sequence of convolutional and pooling layers. The LSTM layer is located just after.
    The right branch (light orange) is dedicated to encoding data matrices. It has an MLP-like structure.
    Each branch returns a latent representation that is conveyed in a concatenation layer.
    A last propagation in a prediction layer (dense layer with sigmoid activation function) returns the desired output.
    }
    \label{fig:TB-ANN_map}
\end{figure}
Finally, both the latent representations are therefore concatenated and then propagated through a dense layer with one output and a sigmoid activation function.
Like other ANNs, the TB-ANN is trained to optimize the binary cross-entropy loss function through the ADAM algorithm. 
An illustration of the TB-ANN is shown in figure \ref{fig:TB-ANN_map}.

As for the models treated in section \ref{icuai:model_selection}, the fine grid-based search was also applied to draw the final model's architecture.
Thus we optimized the parameters featuring both the two branches as well as the dimensionality of the latent representations derived by these two; the search for the optimal hyper-parameters was conducted by maximizing the overall \ac{AUROC} along the domain of landmark times.
That is, the same architecture was used to validate distinct TB-ANNs, each one trained at one distinct landmark time. 
The overall \ac{AUROC}, intended as the \ac{AUROC} averaged by the risk set at different landmark times (see section dedicated in the manuscript), was the quantity we looked at to asses the goodness of the model's architecture.
To compare the TB-ANN with the Deep LM-CR, we adopted similar strategies to those presented in the manuscript.
Thus, we evaluated the TB-ANN predictions' accuracy through the AUROC metric.
As a secondary metric, we also considered the Brier Score.  
As known Brier Score requires prediction probability instead of prediction scores; we made howsoever use of the Platt scaling \cite[]{platt1999probabilistic} to make the TB-ANN output comparabile with a probability.
Similar to the Deep-CR-LM model, the overall metrics are evaluated using the average over the risk sets of each landmark time, and the 95\% confidence intervals are evaluated by means of bootstrap resampling.

\section{Saliency Map Order Equivalent  (SMOE) scale}\label{sec:smoe}
In this section, we present the algorithm used in the manuscript for estimating the saliency maps.
Differently from the gradient-based methods (e.g., the \emph{Vanilla Gradient} (VG), see for instance \cite{simonyan2013deep}), the \ac{SMOE} scale \cite[]{mundhenk2019efficient} provides a different perspective for the estimation of the saliency of the CNN-activated feature maps.
The \ac{SMOE} scale focuses on the statistics of the activation of these feature maps.

The algorithm provides a (reasonably) faithful representation of the information contained in the input data: the larger the overall activation of the feature maps, the more the input features are likely to be informative.
Let us consider a \ac{CNN} model, we denote with $\chi_{ij} \in \mathbb{R}^{+} $ the values of an activated feature map with ReLU activation function and denote with $i$ and $j$, respectively, the spatial domain and the depth (i.e., number of time-series features).
A function $\varphi: \mathbb{R}^{+} \to \mathbb{R}^{+}$ is applied at each point of the spatial domain, all over the depth dimension.
Thus, we obtain the saliency map via the relation $S= \varphi(\chi)$.
We assume that the activated feature map $\chi$ is Gamma distributed with shape parameter $k$ and scale parameter $\theta$.
The reason for this assumption relies on the fact that the Gamma distribution is the \emph{maximum entropy probability distribution} for a random variable whose mean and entropy are fixed (Lagrange multipliers). Since in our context, each activation map has both a fixed mean value (i.e., the scale of the activation map) and fixed entropy (i.e., the information captured in the feature map), the choice of a Gamma distributed feature map seems natural.
Therefore, we estimate the distribution parameters  by means of the Maximum Likelihood Principle, namely:
\begin{equation}
    \label{eq:theta_gamma_best_parameter}
    \hat{\theta}_i = \frac{\sum_{j= 0}^{D} \chi_{ij}}{D\hat{k}_i}, 
\end{equation}
and
\begin{equation*}
    \log{(\hat{k}_i)} - \psi(\hat{k}_i) = \log{\left(\frac{\sum_{j=0}^{D} \chi_{ij}}{D}\right)}-\frac{\sum_{j=0}^{D} \log{\chi_{ij}}}{D};
\end{equation*}
with the sums running over the depth domain (i.e. the domain of the input features), with $D$ the number of input features, and $\psi(x)$ the digamma function \cite[]{silverman1972special}.
We recall that the digamma function is defined as:
\begin{equation*}
    \psi(x) = \frac{d \log{\Gamma(x)}}{dx},
\end{equation*}
with $\Gamma(x)$ the Euler's Gamma function \cite[]{silverman1972special}.

Note that the estimation of the parameters $\theta_i$ and ${k}_i$ is restricted to the $i$-th element along the spatial domain of the activated feature map $\chi$: we are extracting information about the sparseness of activation along the depth domain, and not along the spatial domain.
However, we cannot find an estimation of ${k}_i$ in a closed form, but we can let:
\begin{equation*}
 s_i= \log{\left(\frac{\sum_{j=0}^{D} \chi_{ij}}{D}\right)}-\frac{\sum_{j=0}^{D} \log{\chi_{ij}}}{D};
\end{equation*}
and then we use the asymptotic expansion of the digamma function \cite[]{abramowitz1964handbook}, and we obtain the following approximation:
\begin{equation*}
    \log{x} - \psi({x}) \simeq \frac{1}{2x}\left(1+\frac{1}{6x}\right).
\end{equation*}
Thus, a first order approximation of $\hat{k}$ is given by:
\begin{equation}\label{eq:gamma_params_aprx}
 \hat{k}_i \simeq \frac{\frac{1}{4}+\frac{1}{2}\sqrt{1+3s_i}}{s_i}.
\end{equation}
However,  we can refine $\hat{k}_i$ by using the Newton-Ralphson method \cite[]{ypma1995historical} and use \eqref{eq:gamma_params_aprx} as an initial value.
\begin{figure}
    \centering
    \includegraphics[width=.5\textwidth]{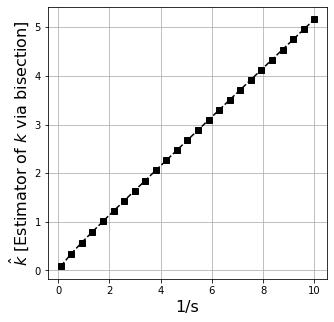}
    \caption{Estimated $\hat{k}$ via Bisection method as a function of the value $1/s$}
    \label{fig:gammapars_k_s}
\end{figure}
As a result, $\hat{k}_i$ appears to be related to $\frac{1}{s_i}$; as shown in Figure~\ref{fig:gammapars_k_s}. 
After substituting $\frac{1}{s_i}$ with $\hat{k}_i$ in \eqref{eq:theta_gamma_best_parameter} we obtain:
\begin{equation*}
    \hat{\theta}_i = \left(\frac{\sum_{j}^{D} \chi_{ij}}{D} \right)\left[\log{\left(\frac{\sum_{j}^{D} \chi_{ij}}{D}\right)}-\sum_{j= 0}^{D}\frac{\log\chi_{ij}}{D} \right], 
\end{equation*}
which  can be finally rewritten as:
\begin{equation}\label{eq:SmoeMap_long}
    \hat{\theta}_{\textnormal{SMOE},i} = \frac{1}{D}\sum_{j= 0}^{D} \left<\chi\right> \log\frac{\left<\chi\right>}{\chi_{ij}}, 
\end{equation}
where 
\begin{equation*}
\left<\chi\right> := \frac{1}{D}\sum_{m= 0}^{D} \chi_{lm}.
\end{equation*}
Hence, \eqref{eq:SmoeMap_long} represents the \emph{SMOE scale}, i.e. the statistics involved in the computation of the saliency maps:
it is proportional to the activated mean value (along depth) via the term $\left<\chi\right>$. Moreover, it depends on the variance, as we can see by performing a Taylor expansion of $\log{\chi_{ij}}$ around $\left<\chi\right>$:
\begin{equation*}
    \log{\left< \chi\right>}-\left< \log{\chi}\right> \simeq  \frac{\left<\chi\ - \left<\chi\right>\right>^{2}}{2\left<\chi^{2}\right>}.
\end{equation*}
The simplification used in the estimation of the Gamma scale parameters is the \ac{SMOE} to the full iterative scale parameter estimation \cite[]{mundhenk2019efficient}.
%

By construction, we can apply the \ac{SMOE} scale only on one single activated feature map; that is, we can only estimate the informative sparseness of each activated feature map independently.
We can then combine them to obtain an overall measurement of saliency at each spatial/temporal location. 


Therefore, the SMOE scale used is the estimated scale parameter of a Gamma distribution. In Section~6 of the main manuscript, we have used this assumption for deriving the saliency map.
However,  we have checked this hypothesis via a multiple Kolmogorov-Smirnov test, with Bonferroni correction:  for each activation map and for a given value of the temporal domain, we have tested whether the values of the feature are gamma distributed with scale parameter $\hat{\theta}_i$ as estimated in (\ref{eq:SmoeMap_long}). We did not reject the null hypothesis with $\alpha=0.05$.

Before concluding this section, we want to present a brief example of the \ac{XAI} methods we have introduced above.
Let us consider the following toy data set for binary classification of time series: the class 0 is generated by:
\begin{equation*}
    X^{(n)}_{0}(t) = \sin{(2\pi[t+\phi_{n}])}, ~ t\in[0, 1];
\end{equation*}
with $ X^{(n)}_{0}(t)$ denoting the $n$-th instance of the class 0, and $\phi_{n} \overset{i.i.d}{\sim} U(-0.125, 0.125)$.
Likewise, for class 1 we set:
\begin{equation*}
    X^{(n)}_{1}(t) = -\sin{(2\pi[t+\phi_{n}])}, ~ t\in[0, 1].
\end{equation*}
\begin{figure}
    \centering
    \includegraphics[width= 10cm]{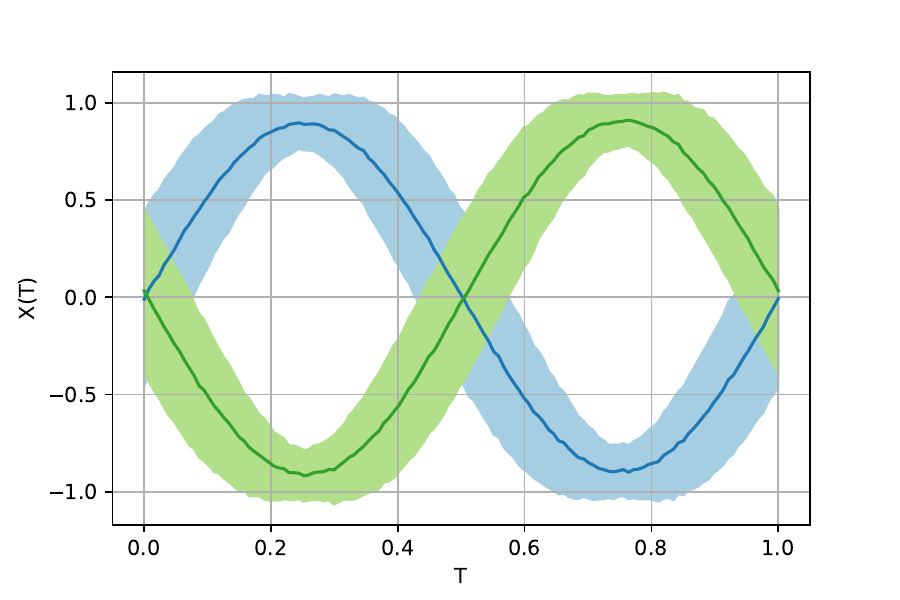}
    \caption{Example of the toy sine dataset. The mean samples are plotted in blue and green, respectively, for classes 0 and 1.
    The light-colored areas represent the standard deviation values of the mean samples. 
    }
    \label{fig:XAI_example_toy_dataset}
\end{figure}
We shall refer to this dataset as \emph{toy sine dataset}.
A representation of the toy sine dataset is shown in Figure~\ref{fig:XAI_example_toy_dataset}.
We train and test a \ac{CNN} with a one-held-out approach (i.e., we only evaluate the model's accuracy after making just one split into train and test set) with train size 75\% (i.e., we use 75\% of the dataset to train the \ac{CNN} model). As a result, the \ac{AUROC} of the model is equal to 0.99.
\begin{figure}
    \centering
    \begin{subfigure}{.39\textwidth}
     \includegraphics[width= \linewidth]{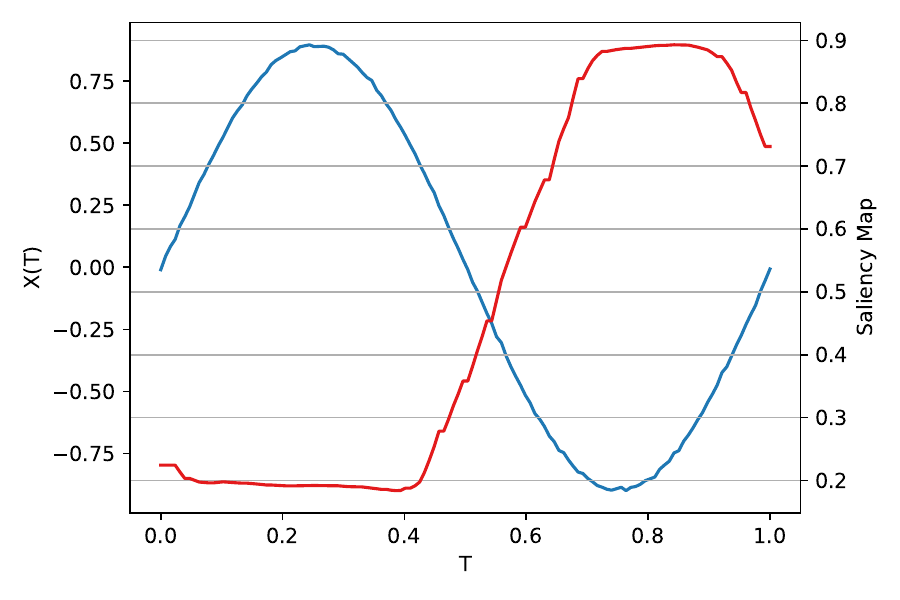}  
     \caption{}
    \end{subfigure}
    \begin{subfigure}{.39\textwidth}
     \includegraphics[width= \linewidth]{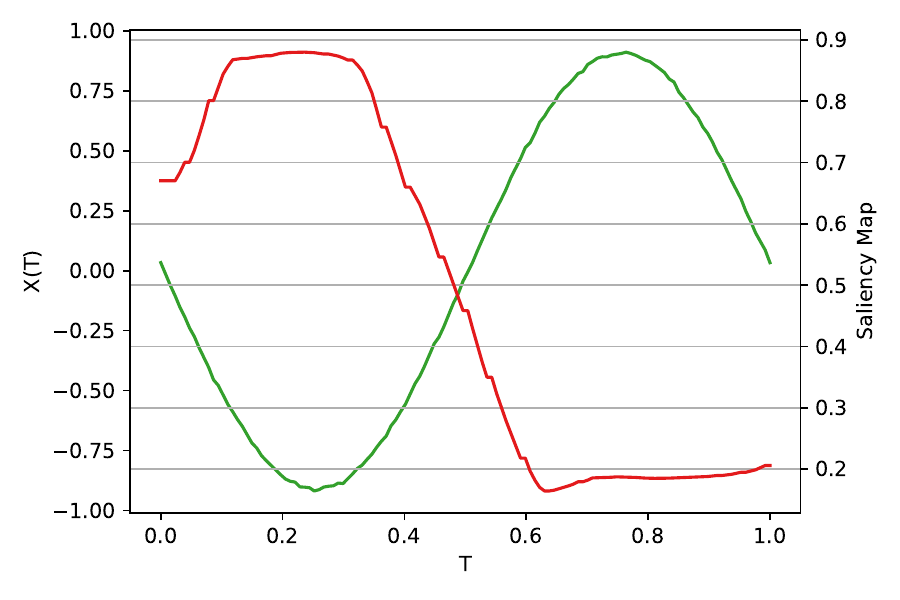} 
    \caption{}
    \end{subfigure}
    \caption{Saliency maps obtained via \ac{SMOE} Scale method for the sine toy dataset. (a) Mean sample (blue) and mean saliency map (red) for class 0.
    (b) Mean sample (blue) and mean saliency map (red) for class 1.
    }
    \label{fig:XAI_example_SMOE}
\end{figure}

Figure~\ref{fig:XAI_example_SMOE} shows the saliency maps for the \ac{SMOE} Scale.
We notice that that the saliency maps detect a salient area in correspondence with the trough of the sinusoidal oscillation, i.e., those areas of the input domain where the saliency maps achieve the highest values.
As expected, the localization of the trough in two distinct areas of the input domain $T\in [0, 1]$ represents the critical feature that the \ac{CNN} captures to distinguish the two classes.

\bibliographystyle{apalike}